\documentclass[fleqn,onecolumn,usedcolumn,usenatbib]{mnras}

\DeclareMathAlphabet{\mathcal}{OMS}{cmsy}{m}{n}
\DeclareMathAlphabet\mathbfcal{OMS}{cmsy}{b}{n}

\usepackage[T1]{fontenc}

\usepackage{graphicx}	
\usepackage{amsmath}	
\usepackage{amssymb}	
\usepackage{bm}
\usepackage{ulem}
\usepackage[dvipsnames]{xcolor}
\usepackage{tablefootnote}

\usepackage{ragged2e}
  
\usepackage{chngpage}
\usepackage{multirow}
\usepackage{tabularx}

\usepackage[utf8]{inputenc}

\newcommand{\euc}{\textit{Euclid}}
\newcommand{\hi}{\textsc{Hi}}
\newcommand{\cosmosis}{\texttt{CosmoSIS}}
\newcommand{\class}{\texttt{CLASS}}
\newcommand{\emcee}{\texttt{emcee}}
\newcommand{\Multinest}{\texttt{Multinest}}
\newcommand{\camb}{\texttt{CAMB}}
\newcommand{\lcdm}{$\Lambda$CDM}
\newcommand{\plk}{$P_{\rm lin}(k)$}
\newcommand{\pk}{$P(k,z)$}
\newcommand{\pgk}{$P_{\rm g}(\bm k,z)$}
\newcommand{\cl}{$C_\ell^{\rm g}$}
\newcommand{\tcl}[2]{$C_\ell^{\rm g}(z_{#1},z_{#2})$}
\newcommand{\tcijl}[2]{$C_\ell^{\rm g}(z_{#1},z_{#2})$}
\newcommand{\de}{\mathrm{d}}
\newcommand{\ho}{H_0}
\newcommand{\om}{\Omega_{\rm m}}
\newcommand{\ob}{\Omega_{\rm b}}
\newcommand{\ns}{n_{\rm s}}
\newcommand{\As}{A_{\rm s}}
\newcommand{\lm}{$\ell_{\rm min}$}
\newcommand{\lM}{$\ell_{\rm max}$}

\title[Unified LSS data analysis pipeline I: RSD in galaxy counts]{Developing a unified pipeline for large-scale structure data analysis with angular power spectra -- I. The importance of redshift-space distortions for galaxy number counts}
\author[K.\ Tanidis \& S.\ Camera]{
Konstantinos Tanidis$^{1,2}$\thanks{E-mail: tanidis@to.infn.it} and Stefano Camera$^{1,2,3,4}$\thanks{E-mail: stefano.camera@unito.it}
\\
$^{1}$Dipartimento di Fisica, Universit\`a degli Studi di Torino, via P.\ Giuria 1, 10125 Torino, Italy\\
$^{2}$INFN -- Istituto Nazionale di Fisica Nucleare, Sezione di Torino, via P.\ Giuria 1, 10125 Torino, Italy\\
$^{3}$INAF -- Istituto Nazionale di Astrofisica, Osservatorio Astrofisico di Torino, strada Osservatorio 20, 10025 Pino Torinese, Italy\\
$^{4}$Department of Physics \& Astronomy, University of the Western Cape, Cape Town 7535, South Africa
}
\date{Accepted XXX. Received YYY; in original form ZZZ}
\pubyear{2019}

\begin{document}
\label{firstpage}
\pagerange{\pageref{firstpage}--\pageref{lastpage}}
\maketitle

\begin{abstract}
We develop a cosmological parameter estimation code for (tomographic) angular power spectra analyses of galaxy number counts, for which we include, for the first time, redshift-space distortions (RSD) in the Limber approximation. This allows for a speed-up in computation time, and we emphasise that only angular scales where the Limber approximation is valid are included in our analysis. Our main result shows that a correct modelling of RSD is crucial not to bias cosmological parameter estimation. This happens not only for spectroscopy-detected galaxies, but even in the case of galaxy surveys with photometric redshift estimates. Moreover, a correct implementation of RSD is especially valuable in alleviating the degeneracy between the amplitude of the underlying matter power spectrum and the galaxy bias. We argue that our findings are particularly relevant for present and planned observational campaigns, such as the \euc\ satellite or the Square Kilometre Array, which aim at studying the cosmic large-scale structure and trace its growth over a wide range of redshifts and scales.
\end{abstract}




\section{Introduction}
\label{sec:int}
The establishment of $\Lambda$ cold dark matter (\lcdm) as the concordance cosmological model has been led by the unprecedented wealth of data obtained over the past decades. Undoubtedly, precise measurements of the cosmic microwave background (CMB) temperature and polarisation anisotropies \citep{Durrer2008, Durrer2015, Ade2013, Ade2015, Ade2015a} have given profound evidence for the validity of this model. However, several analyses and observations show a certain degree of tension among different data sets \citep{Spergel2015,Addison2015,Battye2015,Raveri2016,Joudaki2016a,Joudaki2016b,PourtsidouTram2016,Charnock2017,Camera2017a}. To tackle this issue, and possibly to understand whether these are real hints at the necessity of a change of paradigm in our understanding of the cosmos, a better insight of structure formation and evolution is needed, both on linear and nonlinear scales.

One way to probe the cosmic large-scale structure (LSS) and its growth is by using galaxy catalogues. Galaxy surveys are going to become as powerful as the CMB in constraining cosmological parameters, thanks to the fact that they encode the full three-dimensional (3D) information about the distribution of density fluctuations in the Universe, whereas CMB is ultimately a two-dimensional (2D) surface. Therefore, if we want to study the distribution of galaxies on cosmological scales, we would in principle employ the Fourier-space galaxy power spectrum, \pgk. It is often dubbed `3D' meaning that the wavevector $\bm k$ is the Fourier mode of the 3D separation $\bm s=|\bm x_1-\bm x_2$| between a pair of galaxies located at positions $\bm x_1$ and $\bm x_2$, at redshift $z$. However, to link the galaxy clustering data to the Fourier power spectrum we need to assume a background cosmology. This is due to the fact that what we actually measure is redshifts and angles (or, equivalently, line-of-sight directions $\hat{\bm n}$), meaning that to translate them to 3D positions $\bm x(z,\hat{\bm n})$ we need to assume a cosmological background. Furthermore, the matter power spectrum is a gauge-dependent quantity, and the arbitrariness on the choice of gauge shows up on the largest scales \citep{BonvinDurrer2011,Yoo2010,ChallinorandLewis2011} . On the contrary, the harmonic-space galaxy angular power spectrum, \cl, is a more suitable tool. It represents a natural and gauge-invariant observable for the correlation of galaxy number counts \citep[see e.g.][]{Camera:2018jys}, and it is often referred to as `2D' because it is a summary statistics for the correlation of two sky maps.

Forthcoming galaxy surveys like those that will be performed at optical/near-infrared wavelengths by the European Space Agency \euc\ satellite \citep{Laureijs2011,Amendola2013,Amendola2016}, or in the radio band by the Square Kilometre Array (SKA) \citep{Maartens2015,Abdalla2015,SKA1_2018} will supplement us with information that will push further our knowledge of the Universe. Moreover, synergistic observations at different wavelengths covering large overlapping sky areas will provide us with independent measurements of the clustering and evolution of cosmic structures, thus allowing for valuable cross-correlation studies. This will be a major advantage to tackle systematic effects \citep[see e.g.][]{CameraHarrisonBonaldiBrown}, and possibly to mitigate cosmic variance  \citep{MacDonaldSelijak2009,Selijak2009,FonsecaCameraMaartensSantos}. By doing so, multiple probes will achieve high precision and yield strengthened results on the evaluated cosmological model \citep{Weinberg2013}. Finally, let us emphasise that, besides galaxy clustering, other LSS observables like weak lensing cosmic shear can be employed simultaneously to take better advantage of their complementary information, and to lift degeneracies among cosmological parameters.

A starting point in the literature related to such a synergistic approach has been the combination of the galaxy clustering, galaxy-galaxy lensing and cosmic shear \citep[e.g.][]{Bernstein2009,JoachimiBridle2010,YooSelijak2012,Mandelbaum2013,Cacciato2013,Kwan2016}. Other sophisticated approaches were implemented, e.g.\ \citet{Liu2016} used cross-correlations of CMB lensing with galaxy overdensity and cross-correlations of galaxy overdensity and the shear field to probe the multiplicative bias for CFHTLenS. Such approaches are currently being extensively employed by the Dark Energy Survey Collaboration, \citep[see e.g.][]{Elvin-Poole:2017xsf,Abbott:2018ydy,Abbott:2017wau}. Furthermore, there have been thorough theoretical investigations using non-Gaussian covariances between galaxy clustering,  weak lensing, galaxy-galaxy lensing, galaxy cluster number counts, galaxy clusters and photometric baryon-acoustic oscillations for photometric galaxies \citep{Eifler2014,KrauseEifler2016}, also with the inclusion of CMB data \citep{NicolaRefregierAmara2016,Singh2016}.

Within such a wider context, our present paper is the first of a series in which we aim to go beyond standard Fisher matrix analyses for the tomographic angular power spectrum of galaxy number counts. Here, we focus only on forecasts for single probes using galaxy clustering, and leave other observables, their cross-correlation, and multi-tracing for future works. We consider two broad families of galaxy surveys, both of which are used to probe the cosmic LSS. One of them is represented by the spectroscopic observations, where the redshift of the galaxies is inferred with high accuracy. The other deals with photometric surveys, where galaxies are binned into broad-band redshift slices, due to the large uncertainty in the determination of photometric redshifts. A noteworthy work is that of \citep{Jonas2018} where they studied the effect of photo-z errors on the galaxy number counts using the Fourier-space power spectrum.  We, on the other hand, aim to study galaxy number counts by measuring the tomographic angular power spectrum, \tcl{i}{j}, in different redshift bins, $z_i$ and $z_j$. The importance of the tomographic approach in galaxy clustering using the density fluctuations with auto- and cross-spectra between photometric redshift bins, has been studied by \citep{Balaguera-Antolinez2018} with the 2MPZ catalogue at the local universe. To this purpose, we adopt as proxies of the two aforementioned families of galaxy surveys a \euc-like photometric instrument and the specifications of \hi-line galaxy observations with the Phase 1 of the SKA (SKA1). We perform an extensive Bayesian analysis for the two showcases, for which we generate synthetic data including both leading-order Newtonian density fluctuations and the linear-order contribution due to redshift-space distortions (RSD) \citep[e.g.][]{Kaiser1987,Szalay1998}. Some original pieces of work which considered a spherical harmonic analysis in redshift space are \citep{Fisher1993,Heaven1995}. In particular, we provide the reader with an expression for RSD in Limber approximation \citep{Limber1953,LoverdeAfshordi2008}. To our knowledge, this is the first in the literature. 
The paper is organised as follows. In \autoref{sec:APS} we introduce the tomographic angular power spectrum \tcijl{i}{j} with and without RSD \citep{Limber1953,Kaiser1992}, which we implement in the public \cosmosis\ code \citep{Zuntz2015} by using today's Fourier-space linear power spectrum \plk\ provided by \camb\ \citep{LCL2000}. A comparison between our Limber approximated spectra obtained with our modified \cosmosis\ module and the full solution provided by \class\ \citep{Lesgourgues2011,Blas2011,DiDio2013} is presented in \autoref{sec:comparison} for different test window functions. In \autoref{sec:survey} we present the surveys specifications and then in \autoref{sec:implementation}, we compare the equi-spaced and equi-populated binning scenarios via Fisher matrices for an idealistic case involving cosmological parameters only. In addition we show the likelihood applied in the final analysis. In \autoref{sec:res}, we perform the Bayesian forecasting analysis for the same idealistic case and then including real-world nuisance parameters. Drawn conclusions are discussed in \autoref{sec:disc}. 

Throughout the paper, we assume a fiducial \lcdm\ model with the best-fit parameters as of \citet{Ade2015} (see \autoref{tab:params} in \autoref{sec:res} for symbols and fiducial values).

\section{The angular power spectrum of galaxy number counts}
\label{sec:APS}
Here, we introduce the main tool of our analysis, i.e.\ the tomographic angular power spectrum of galaxy number counts in the Limber approximation, for which we include RSD for the first time. To do so, we start from the Fourier-space matter power spectrum, \pk, and at the end apply the Limber approximation to the harmonic-space angular power spectrum, \tcl{i}{j}. We modify modules of the publicly available \cosmosis\ code. We check the agreement between our approximated spectra and the full solution provided by the \class\ Boltzmann solver (see ~\autoref{sec:comparison}).

\subsection{The Fourier-space matter power spectrum}
\label{sec:Pk}
The linear matter power spectrum is 
\begin{align}
P_{\rm lin}(k,z)&=
\frac{8\upi^2}{25\ho^4\om^2g_\infty^2}D^2(z)T^2(k)\mathcal P_\zeta(k)k\nonumber\\
&=P_{\rm lin}(k)D^2(z),
\label{eq:matter}
\end{align}
where $\ho$ is the Hubble constant today, $\om$ the fractional matter density, and we have exploited the fact that, in general relativity and in the absence of anisotropic stress, we can separate scale and redshift dependence thus having a redshift-independent transfer function, $T(k)$, and a scale-independent growth factor, $D(z)$; here, $g_\infty=\lim_{z\to\infty}(1+z)D(z)\simeq1.27$. $\mathcal P_\zeta(k)=\As(k/k_0)^{\ns-1}$ is the dimensionless power spectrum of the primordial curvature perturbation. We also define the present-day linear matter power spectrum as $P_{\rm lin}(k)\equiv P_{\rm lin}(k,z=0)$. Hereafter, we shall limit our analysis to linear scales.

\subsection{The harmonic-space galaxy angular power spectrum}
\label{sec:Cl}
On linear scales, it is customary to define the (tomographic) angular power spectrum of a generic observable $X$ as
\begin{equation}
C^X_{\ell}(z_i,z_j)=4\upi\int\de\ln k\,\mathcal P_\zeta(k)\mathcal W_\ell^X(k;z_i)\mathcal W_\ell^X(k;z_j),
\label{eq:Cl_fullsky}
\end{equation}
with $\mathcal W_\ell^X(k;z_i)$ denoting the weight function for observable $X$ in the $i$th redshift bin. In the case of galaxy number counts (i.e.\ $X={\rm g}$), the weight function reads
\begin{equation}
\mathcal W^{\rm g}_\ell(k;z_i)=\int\de\chi\,n^i(\chi)\mathcal W^{\rm g}_\ell(k,\chi),\label{eq:weight_func}
\end{equation}
where $\chi=\chi(z)$ is the radial comoving distance to redshift $z$, and $n^i(\chi)$ is the redshift distribution of sources in bin $i$, for which both $n^i(\chi)\de\chi=n^i(z)\de z$ and $\int\de z\,n^i(z)=1$ hold. In longitudinal gauge, and including up to RSD, we have
\begin{equation}
\mathcal W^{\rm g}_\ell(k,\chi)=b(\chi)D(\chi)T(k)j_\ell(k\chi)-f(\chi)D(\chi)T(k)j^{\prime\prime}_\ell(k\chi),\label{eq:den+RSD}
\end{equation}
with $b$ the linear galaxy bias, $f\equiv-(1+z)\de\ln D/\de z$ the growth rate, and $j_\ell$ the spherical Bessel function of order $\ell$. (A prime denotes derivatives with respect to the argument of the function, viz.\ $k\chi$.) The first term in \autoref{eq:den+RSD} is the main contribution to galaxy number density fluctuations, due to density perturbations, whereas the second term encodes RSD.

The computation of angular power spectra as in \autoref{eq:Cl_fullsky} is time expensive and prone to numerical instabilities, due to the integration of highly oscillating spherical Bessel functions. Therefore, the Limber approximation (valid on scales $\ell\gg1$) is often employed. In this limit, the spherical Bessel functions are proportional to a  Dirac Delta,
\begin{equation}
j_\ell(k\chi)\underset{\ell\gg1}{\longrightarrow}\sqrt{\frac{\upi}{2\ell+1}}\delta_{\rm D}\left(\ell+\frac{1}{2}-k\chi\right).
\end{equation}
By inserting this into \autoref{eq:Cl_fullsky}, and for now just considering the first term in \autoref{eq:den+RSD}, we obtain the well-known expression for the galaxy angular power spectrum in Limber approximation,\footnote{Henceforth, we shall use, in comparisons, `den+RSD' and `den' to refer to \autoref{eq:CldenRSD_Limber} or \autoref{eq:Clden_Limber}, respectively. Otherwise, when no ambiguity arises, \tcl{i}{j} will either mean the galaxy angular power spectrum in general, or the most comprehensive case considered in this paper, viz.\ `den+RSD'.}
\begin{equation}
C^{\rm g,den}_{\ell\gg1}(z_i,z_j)=\int\de\chi\,\frac{W_b^i(\chi)W_b^j(\chi)}{\chi^2}P_{\rm lin}\left(k=\frac{\ell+1/2}{\chi}\right).
\label{eq:Clden_Limber}
\end{equation}
Since the contribution to galaxy number counts from density fluctuations is modulated by the galaxy bias, we have defined the window function
\begin{equation}
W_b^i(\chi)=n^i(\chi)b(\chi)D(\chi).
\label{eq:W_den}
\end{equation}

Now, we want to include RSD in the Limber galaxy angular power spectrum. As clear from \autoref{eq:den+RSD}, RSD are driven by the growth rate, $f(z)$, we thus introduce a new window function,
\begin{equation}
W_f^i(\chi)=n^i(\chi)f(\chi)D(\chi).
\label{eq:W_RSD}
\end{equation}
After some manipulations (see \autoref{sec:apa1}), and the introduction of a window function for the global `den+RSD' signal,
\begin{equation}
W^i(\chi)=
W_b^i(\chi)+\frac{2\ell^2+2\ell-1}{(2\ell-1)(2\ell+3)}W_f^i(\chi)-\frac{(\ell-1)\ell}{(2\ell-1)\sqrt{(2\ell-3)(2\ell+1)}}W_f^i\left(\frac{2\ell-3}{2\ell+1}\chi\right)-\frac{(\ell+1)(\ell+2)}{(2\ell+3)\sqrt{(2\ell+1)(2\ell+5)}}W_f^i\left(\frac{2\ell+5}{2\ell+1}\chi\right),
\label{eq:W_tot}
\end{equation}
we eventually get
\begin{equation}
C^{\rm g,den+RSD}_{\ell\gg1}(z_i,z_j)=\int\de\chi\,\frac{W^i(\chi)W^j(\chi)}{\chi^2}P_{\rm lin}\left(k=\frac{\ell+1/2}{\chi}\right).
\label{eq:CldenRSD_Limber}
\end{equation}

It is instructive to notice how RSD affect the harmonic-space angular power spectrum. It is known that the Fourier-space galaxy power spectrum \pgk, which is isotropic
if we consider density fluctuations only, due to RSD acquires a further dependence on $\mu$, the cosine between the wave-vector ${\bm k}$ and the line-of-sight direction $\hat{\bm n}$. This translates into a quadrupolar anisotropy pattern, resulting into the well-known squashing of the galaxy power spectrum on large scales and in the direction perpendicular to the line of sight, and, oppositely, into the so-called Finger-of-God effect on nonlinear scales and in the line-of-sight direction. On the contrary, the net effect of RSD on the harmonic-space angular power spectrum \cl\ is far less straightforward. In this sense, the Limber approximation makes it simpler to understand. If we look at \autoref{eq:W_tot}, we appreciate that RSD effectively shuffle galaxies around among (neighbouring) redshift bins due to the $(2\ell-3)/(2\ell+1)$ and $(2\ell+5)/(2\ell+1)$ factors that modulate $\chi$ in the RSD window functions. The reason behind this is the second derivative of the spherical Bessel function in \autoref{eq:den+RSD}, in turn coming from RSD being caused by the radial derivative of the galaxies' velocity along the line of sight \citep[see e.g.][Section~III]{BonvinDurrer2011}. As in the case of the Fourier-space galaxy power spectrum discussed above, linear RSD effects are stronger on the largest angular scales, where $(2\ell-3)/(2\ell+1)$ or $(2\ell+5)/(2\ell+1)$ deviate from unity the most. (We remind the reader that we limit our analysis to linear scales, so we are not interested in modelling Finger-of-God effects.)

\section{Surveys adopted in the analysis}
\label{sec:survey}
Here, we present the details of the two surveys adopted to test our pipeline. One survey is a proxy for future photometric imaging experiments, and the other is a representative of planned spectroscopic observational campaigns. Better to foresee the potentiality of our pipeline when applied to oncoming data from cosmological galaxy surveys, we decide to study both the cases of optical/near-infrared and radio observations.

To model redshift binning in spectroscopic and photometric redshift surveys, we here assume top-hat and Gaussian bins, respectively. This is clearly a simplification, but it is enough to capture the main features of both observational strategies. On the one hand, the exquisite redshift accuracy of spectroscopic measurements allows for separating galaxies into sharp, non-overlapping redshift slices. This is implemented here by the top-hat bins, to which we had a degree of smoothing to stabilise numerical integration over the bin. On the other hand, photometric redshift estimation is far less accurate than spectroscopy, and it usually results into a PDF $p(z_{\rm ph}|z)$ for each galaxy, representing the probability of having estimated a photometric redshift, $z_{\rm ph}$, given the galaxy's true redshift, $z$. Although one could, in principle, use each galaxy separately \citep[see e.g.][]{Kitching:2010wa}, it is customary to combine the various PDFs into a certain number of redshift bins, which look much broader than spectroscopic ones, and which often overlap each other to a greater or lesser extent, depending on photometric redshift uncertainties. Without any loss of generality, we follow the literature and model this effect by implementing Gaussian redshift bins with a redshift-dependent (monotonically-increasing) width.

For a generic survey $X$, we shall denote: the total redshift distribution of sources by $n_X(z)$; the distribution of sources in the $i$th redshift bin by $n_X^i(z)$; and the (angular) number density of galaxies by
\begin{equation}
\bar n_X^i=\int\de z\,n_X^i(z),\label{eq:n_i}
\end{equation}
so that the total number density of galaxies is $\bar n_X=\sum_i{\bar n_X^i}$.\footnote{We remind the reader that the term $n^i(z)$ appearing in \autoref{eq:weight_func}, \autoref{eq:W_den}, and \autoref{eq:W_RSD} is normalised, meaning that it in fact corresponds to $n_X^i(z)/\bar n_X^i$.} The redshift distributions for the two surveys under investigation, and the two binning strategies are shown in \autoref{fig:d}, and will be discussed in the following sections.

\subsection{Photometric galaxy survey}\label{sec:e1}
\begin{figure*}
\centering
\includegraphics[width=0.45\textwidth]{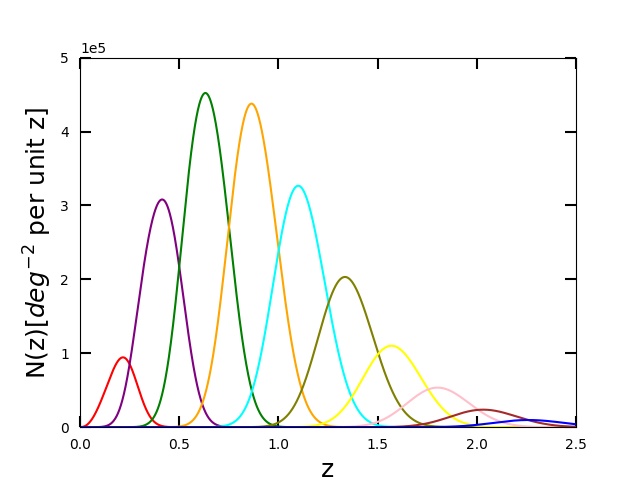}\includegraphics[width=0.45\textwidth]{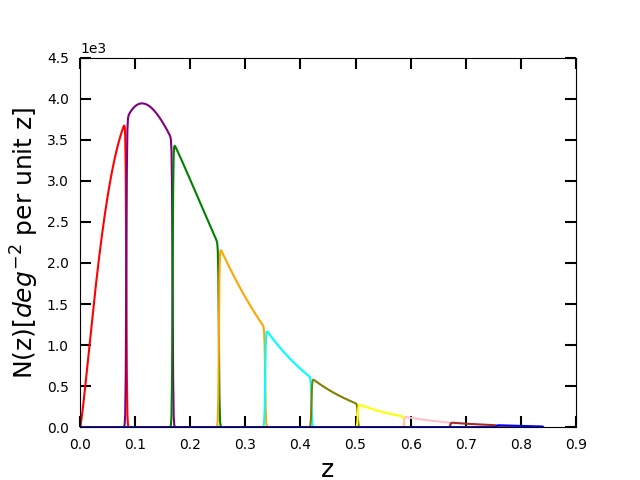}\\\includegraphics[width=0.45\textwidth]{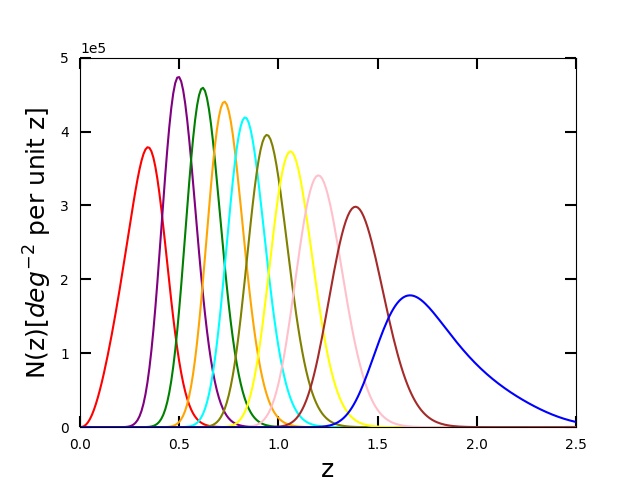}\includegraphics[width=0.45\textwidth]{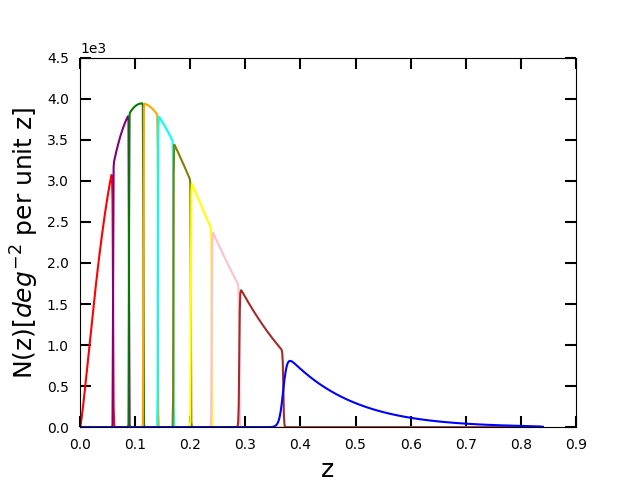}
\caption{Galaxy redshift distributions for the \euc-like photometric galaxy survey (left panels) and the SKA1 \hi-line galaxy survey (right panels). Top and bottom panels respectively show equi-spaced and equi-populated bins.}
\label{fig:d}
\end{figure*}

As a proxy of an optical/near-infrared photometric galaxy survey, we adopt the specifications of a \euc-like experiment \citep{Laureijs2011,Amendola2013,Amendola2016}. The \euc\ satellite will be launched in 2021 and will probe $15,000\,\mathrm{deg}^2$ of the sky for weak lensing and photometric galaxy clustering in the redshift range $0<z\lesssim2.5$, detecting $\bar n_{\rm Euc}=30$ galaxies per square arcminute. The source redshift distribution and the redshift-dependent galaxy bias are given by \citep{Laureijs2011}
\begin{align}
n_{\rm Euc}(z)&=\frac{3\bar n_{\rm Euc}}{2z_0^3}z^2{\exp}{\left[{-\left(\frac{z}{z_0}\right)^{3/2}}\right]}\,\mathrm{arcmin}^{-2},\\
b_{\rm Euc}(z)&=\alpha_{\rm Euc}(1+z)^{\beta_{\rm Euc}}\label{eq:eucb},
\end{align}
where $z_0=0.9/\sqrt{2}$, $0.9$ being the mean redshift of the survey, $\alpha_{\rm Euc}=1$, and $\beta_{\rm Euc}=0.5$. In \autoref{fig:d} (left panels) we present the equi-spaced and equi-populated binned $n_{\rm Euc}(z)$, implementing photometric redshift errors. We use photometric uncertainties in redshift following \citet{MaHuHuterer2006}. That is, the given true redshift distribution of galaxies inside the $i$th photometric redshift bin with photometric redshift estimate $z_{\rm ph}$ in the range $z_{\rm ph}^i<z_{\rm ph}<z_{\rm ph}^{i+1}$ can be expressed as 
\begin{equation}
n_{\rm Euc}^i(z)=\int_{z_{\rm ph}^i}^{z_{\rm ph}^{i+1}}\!\!\de z_{\rm ph}\,n_{\rm Euc}(z)p(z_{\rm ph}|z),
\end{equation}
where $p(z_{\rm ph}|z)$ is the probability distribution of photometric redshift estimates $z_{\rm ph}$ given true redshifts $z$. More specifically, we adopt a probability distribution of Gaussian form,
\begin{equation}
p(z_{\rm ph}|z)=\frac{1}{{\sqrt{2\upi}}\sigma_z}\exp\left[-\frac{\left(z-z_{\rm ph}-\delta z\right)^2}{2\sigma_z^2}\right],
\end{equation}
with $\delta z$ the redshift bias (set to zero in our case), and $\sigma_z=$0.05$(1+z)$ the scatter of the photometric redshift estimate with respect to the true redshift value--- a typical value in photometric redshift measurements \citep[see e.g.][]{Hoyle:2017mee}.

\subsection{Spectroscopic galaxy survey}
\label{sec:e2}
As a representative of oncoming cosmological experiments operating at radio frequencies, we choose a spectroscopic \hi\ galaxy survey performed by SKA1 \citep[][]{Maartens2015,Abdalla2015,SKA1_2018}, which will be able to access even very large angular scales \citep{Camera:2014bwa,CMS2014}. Such a survey with this large radio telescope will probe $5000\,\mathrm{deg}^2$, detecting $\bar n_{\rm SKA}=0.28$  galaxies per square arcminute \citep[][`reference' case]{Yahya}. The survey specifications adopted in this paper for the range $0<z\lesssim0.9$ are
\begin{align}
n_{\rm SKA}(z)&= 10^{5.438}\,z^{1.332}\,{\rm e}^{-11.837z}\,\mathrm{deg}^{-2},\label{eq:ska}\\
b_{\rm SKA}(z)&=\alpha_{\rm SKA}\exp\left(\beta_{\rm SKA}z\right)\label{eq:skaa},
\end{align}
with $\alpha_{\rm SKA}=0.625$ and $\beta_{\rm SKA}=0.881$. Similarly to the case of \euc, we consider equi-spaced and equi-populated bins as shown in \autoref{fig:d} (right panels). In both scenarios we choose 10 bins. For the top-hat bins, we define a smoothed top-hat window function (the same functional form is implemented in \class), i.e.\
\begin{equation}
w_{\rm SKA}(z)=\frac{1}{2}\left\{1-\tanh\left[\frac{|z-\bar z|-\sigma_{\rm SKA1}}{r\sigma_{\rm SKA}}\right]\right\},
\label{eq:edgee}
\end{equation}
where $\bar z$ is the central value of the bin, $\sigma_{\rm SKA}$ is half of the top-hat width, and $r$ is the smoothing edge factor, with a realist value of 0.03.

\section{Pipeline implementation}\label{sec:implementation}
Here we describe the various ingredients and tests performed to implement and validate our pipeline.

\subsection{Validation of the code}
\label{sec:comparison} Here, we perform some tests to validate the expressions derived in \autoref{sec:Cl}, namely the agreement between the Limber approximation in \autoref{eq:CldenRSD_Limber} and the full solution involving the double integral and the spherical Bessel functions of \autoref{eq:Cl_fullsky}. We consider four window functions for the angular power spectrum. Our code is validated against the results of \class, where the Limber approximation is applied for multipoles $\ell\geq100$, but we also cross-checked that our results do not change if we enforce \class\ never to use the Limber approximation.

Consequently, these cases are considered to be indicative of the binning scenarios for \euc\ and SKA1 as shown in \autoref{sec:e1} and \autoref{sec:e2} and are chosen as templates to validate the performance of the code.

\begin{figure*}
\centering
\includegraphics[width=0.45\textwidth]{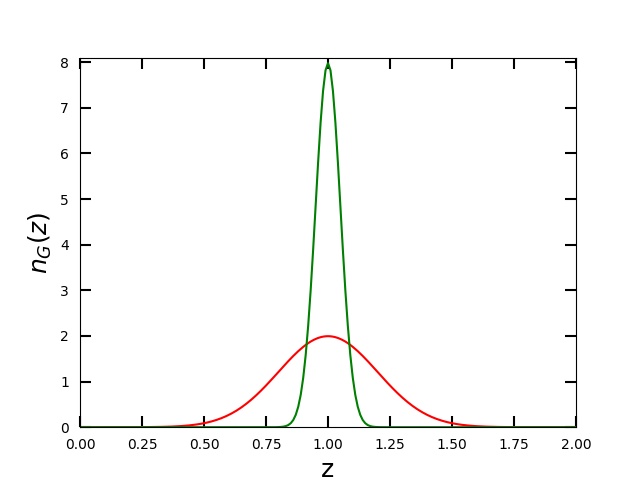}\includegraphics[width=0.45\textwidth]{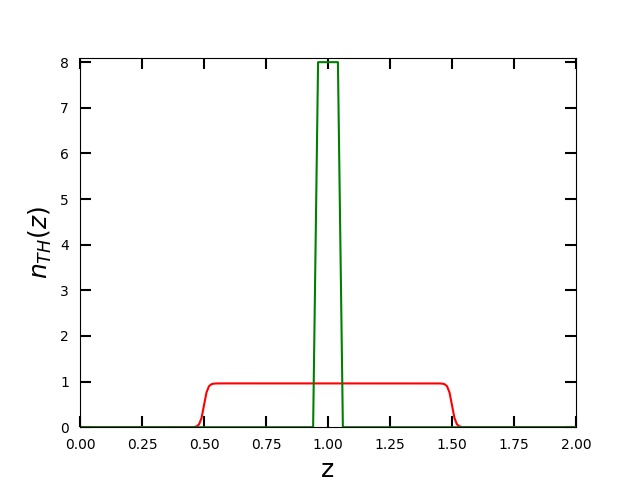}
\caption{Various window functions: broad and narrow Gaussian (left panel), and broad and narrow smoothed top-hat (right panel).}\label{fig:GaussTH}
\end{figure*}

For the sake of simplicity, let us assume that we have only one redshift bin covering the range $0<z\leq2$ and peaking at $\bar z=1$. We can define a Gaussian distribution of sources in the bin as 
\begin{equation}
n_{\rm G}(z)=\frac{1}{\sigma_{\rm G}{\sqrt{2\upi}}}\exp\left[{-\frac{(z-\bar z)^2}{2\sigma_{\rm G}^2}}\right],
\label{eq:gaussy}
\end{equation}
where $\sigma_{\rm G}$ is the width of the distribution. We consider both a narrow and a broad bin by setting $\sigma_{\rm G}=0.05$ and $\sigma_{\rm G}=0.2$, respectively. Such a Gaussian bin is shown in the left panel of \autoref{fig:GaussTH}.

Similarly, we adopt Eq.~\ref{eq:edgee} where now $\sigma_{\rm TH}$ is half of the top-hat width, and $r$ is the smoothing edge factor. Again, we consider both a narrow and a broad redshift bin, respectively defined by $\{\sigma_{\rm TH},r\}=\{0.05,0.003\}$ and $\{0.5,0.03\}$. They are presented in the right panel of \autoref{fig:GaussTH}.

\begin{figure*}
\hspace*{-.20\textwidth}%
\begin{minipage}{17cm}
\includegraphics[width=0.65\textwidth]{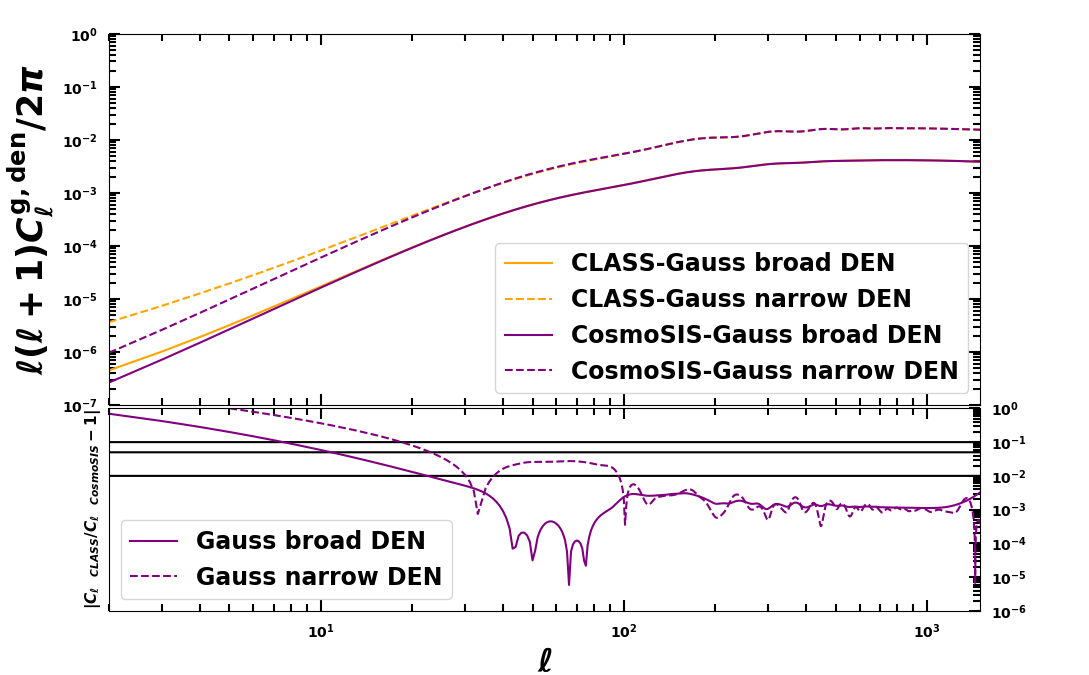}\hspace*{-.06\textwidth}%
\includegraphics[width=0.65 \textwidth]{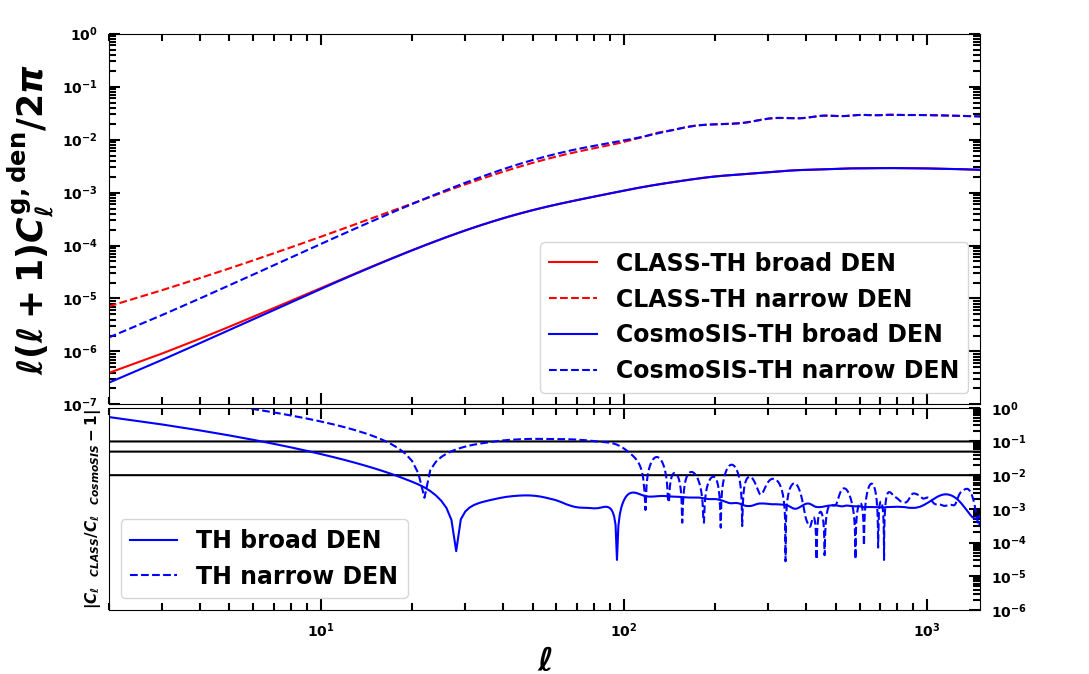}\\
\includegraphics[width=0.65\textwidth]{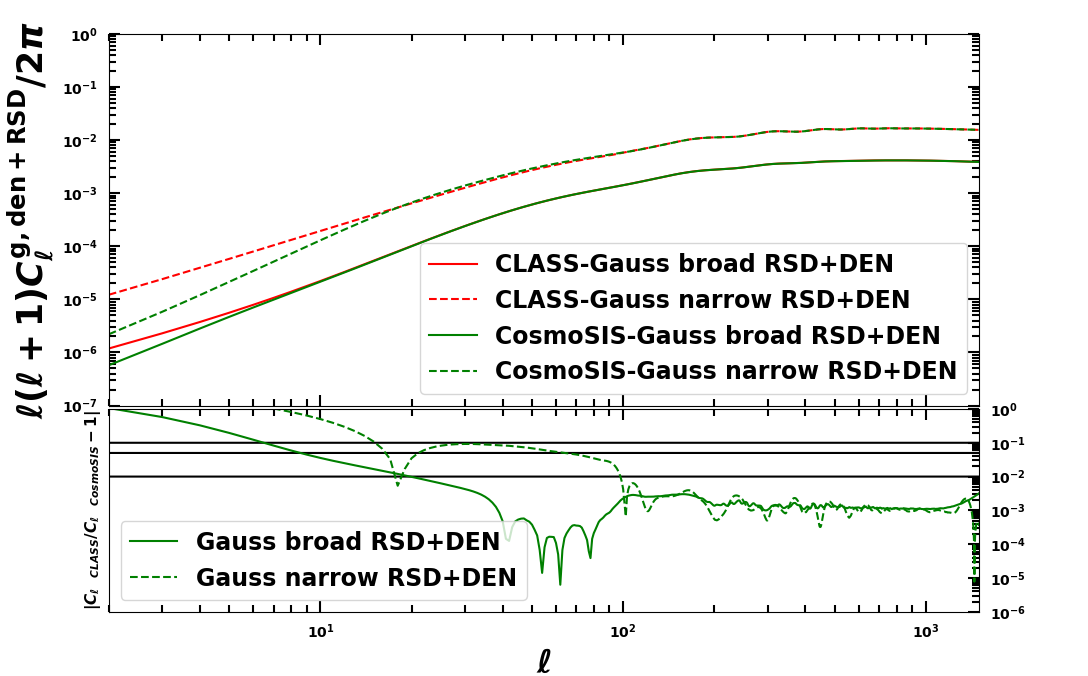}
\hspace*{-.064\textwidth}%
\includegraphics[width=0.65\textwidth]{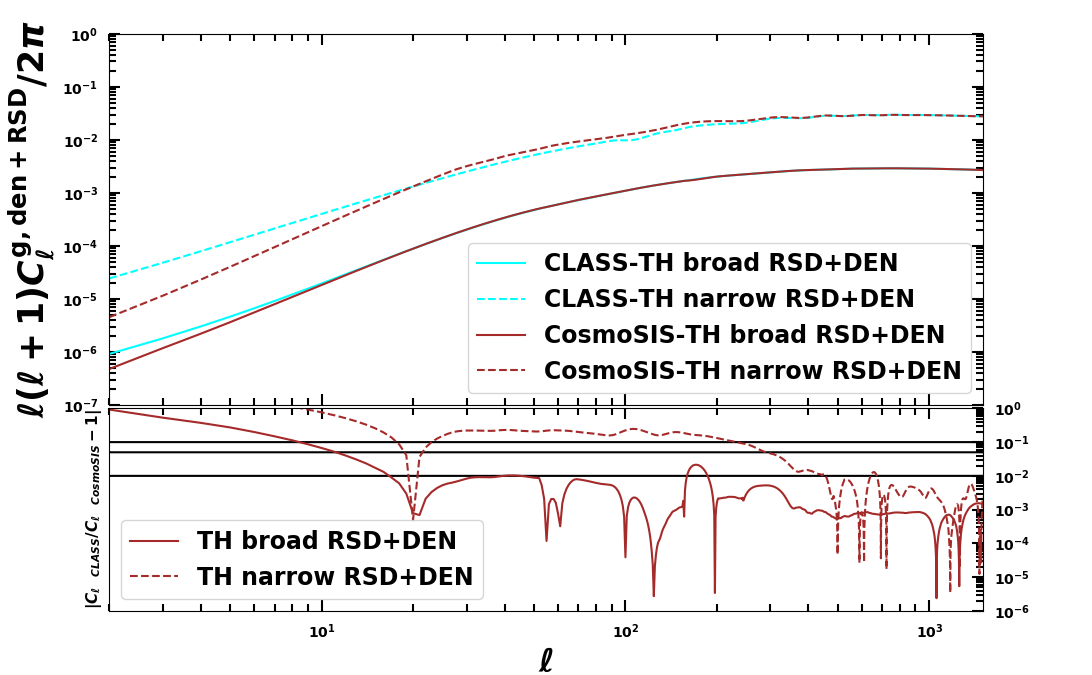}
\end{minipage}
\caption{Code comparison for the window functions (solid lines: broad bins; dashed lines: narrow bins). Top and bottom panels respectively refer to `den' and `den+RSD', with Gaussian (top-hat) window functions on the left (right). In each panel, the bottom plot shows the relative error due to Limber approximation as implemented in our modified version of \cosmosis, with respect to the full solution of \class; the three black solid lines correspond to $10\%$, $5\%$ and $1\%$ relative errors from top to bottom, respectively.}\label{fig:d2}
\end{figure*}

We check our code performance against the \class\ for the case of density perturbations only in \autoref{fig:d2} (top panels) for the broad and narrow Gaussian and top-hat bins. Similarly, the convergence is shown for the case of density and RSD as seen in \autoref{fig:d2} (bottom panels).

\subsection{Multipole range}
Since the Limber approximation is not a good approximation on large angular scales, we set the minimum multipole in our analysis, \lm, by performing the same comparison as in \autoref{fig:d2} for each bin pair, binning scenario, and survey. For the rest of this analysis, we consider the convergence between Limber-approximated spectra and the full solution of \autoref{eq:Cl_fullsky} met when the relative error between \cosmosis\ and \class\ is below 5\%. This is a reasonable choice, since such a percentage difference between correct and approximated angular power spectra is well within the standard deviation of the signal (see \autoref{sec:like} for the covariance matrix). The result of this is presented in \autoref{tab:multipoles}.
\begin{table*}
\centering
\caption{Mininum and maximum multipoles for the two binning strategies. The former are set so that the relative error between angular spectra computed with \cosmosis\ and \class\ is below 5\%. The latter follow $\ell_{\rm max}=\chi(\bar z_i)k_{\rm max}$ in redshift bin $i$ centred on $\bar z_i$.}
\begin{tabular}{ccccccccccccc}
\hline
\multicolumn{6}{c}{Equi-spaced bins} && \multicolumn{6}{c}{Equi-populated bins} \\
\cline{1-6}\cline{8-13}
\multicolumn{3}{c}{\euc} & \multicolumn{3}{c}{SKA1} && \multicolumn{3}{c}{\euc} & \multicolumn{3}{c}{SKA1} \\
\cline{1-6}\cline{8-13}
\multicolumn{2}{c}{\lm} & \multicolumn{1}{c}{\lM} & \multicolumn{2}{c}{\lm} & \multicolumn{1}{c}{\lM} && \multicolumn{2}{c}{\lm} & \multicolumn{1}{c}{\lM} & \multicolumn{2}{c}{\lm} & \multicolumn{1}{c}{\lM} \\
\cline{1-2}\cline{4-5}\cline{8-9}\cline{11-12}
den & den+RSD & & den & den+RSD & && den & den+RSD & & den & den+RSD & \\
\hline
\hline
2 & 2 & 133 & 3 & 13 & 45 && 4 & 3 & 348 & 2 & 7 & 32 \\

8 & 6 & 373 & 1 & 13 & 134 && 10 & 7 & 480 & 7 & 30 & 80 \\

12 & 9 & 581 & 14 & 26 & 218 && 12 & 9 & 576 & 11 & 78 & 109 \\

16 & 11 & 759 & 29 & 40 & 299 && 15 & 10 & 659 & 13 & 77 & 136 \\

22 & 13 & 913 & 33 & 60 & 375 && 17 & 12 & 733 & 15 & 80 & 164 \\

28 & 17 & 1046 & 43 & 70 & 448 && 18 & 13 & 806 & 19 & 80 & 194 \\

32 & 20 & 1162 & 63 & 73 & 518 && 20 & 14 & 880 & 22 & 91 & 228 \\

36 & 22 & 1265 & 60 & 101 & 584 && 22 & 15 & 957 & 26 & 82 & 270 \\

40 & 25 & 1356 & 70 & 110 & 647 && 24 & 17 & 1054 & 30 & 65 & 331 \\

50 & 30 & 1437 & 80 & 120 & 707 && 25 & 19 & 1181 & 11 & 44 & 564 \\
\hline
\end{tabular}
\label{tab:multipoles}
\end{table*}

Generally, it is evident that there is a trend of increasing \lm\ with redshift, apart from the equi-populated bins for SKA1, to whose highest $z$ bin(s) correspond a lower \lm. This happens because the broader the top-hat bin, the more accurate the Limber approximation (see also the right panels of Fig.~\ref{fig:d2}). Interestingly, we also find that in the case of the smoother, photometric redshift bins of the \euc-like survey, the agreement between Limber and non-Limber spectra extends to larger scales when RSD are included, than what happens with density perturbations only.

Additionally, we want to find the upper limits of the multipole range for each redshift bin so that we safely remain within the linear regime. This corresponds to setting the largest angular scale, \lM, corresponding to the maximum wavenumber before entering the nonlinear regime, $k_{\rm max}$. This is estimated through the rms fluctuations of the total mass density in spheres of radius $R$ at $z=0$, 
\begin{equation}
\sigma_M^2(R)=\int\frac{\de k}{2\upi^2}\,k^2P_{\rm lin}(k)\left[\frac{3j_1(kR)}{kR}\right]^2.
\label{eq:highmul}
\end{equation}
We choose $k_{\rm max}$ such that $\sigma_M^2(R_{\rm min})=1$ and $k_{\rm max}=\upi/(2R_{\rm min})$. Since we are considering multipoles $\ell\gg1$, where the Limber approximation is a good approximation, we simply set $\ell_{\rm max}=k_{\rm max}\chi(\bar z_i)$, with $\bar z_i$ the centre of the $i$th redshift bin. We find $k_{\rm max}=0.2469\,h\,\mathrm{Mpc}^{-1}$ for our fiducial model.

\subsection{Likelihood}
\label{sec:like}
To construct the likelihood of the signal, we start from the Gaussian covariance matrix implemented in \cosmosis, $\boldsymbol\Gamma_{\ell\ell^\prime}$, whose entries are
\begin{equation}
\Gamma_{\ell\ell^\prime}^{ij,kl} = \frac{\delta_{\rm K}^{\ell\ell^\prime}}{2\ell\Delta\ell f_{\rm sky}}\left[\widetilde C^{\rm g}_\ell(z_i,z_k)\widetilde C^{\rm g}_\ell(z_j,z_l)+\widetilde C^{\rm g}_\ell(z_i,z_l) \widetilde C^{\rm g}_\ell(z_j,z_k)\right],\label{eq:covmat}
\end{equation}
where $\Delta\ell$ is the width of the multipole bin, $f_{\rm sky}$ the sky fraction covered by the survey, $\delta_{\rm K}$ is the Kronecker symbol, and the observed signal is
\begin{equation}
\widetilde C^{\rm g}_\ell(z_i,z_j)=C^{\rm g}_\ell(z_i,z_j)+\frac{\delta_{\rm K}^{ij}}{\bar n^i},
\end{equation}
with $\bar n^i$ defined in \autoref{eq:n_i}.\footnote{Note that the denominator of \autoref{eq:covmat} should actually read $(2\ell+1)$, and not $2\ell$ as reported in \citet{JoachimiBridle2010}. Such a difference, however, is negligible for $\ell\gg1$ where the Limber approximation holds true. Moreover, we are here interested in comparing two methods (i.e.\ fitting the data with or without RSD), so the absence of the $+1$ factor does not affect the validity of our results.} Then, for $N_z$ redshift bins and $N_\ell=20$ multipole bins, we write the data vector as
\begin{equation}
\bm d_\ell=\bigg\{\underbrace{ C^{\rm g}_{\ell_{\rm min}}(z_1,z_1),\ldots, C^{\rm g}_{\ell_{\rm min}}(z_1,z_{N_z})}_{\begin{minipage}{0.2\textwidth}\small{Auto- and cross-bin spectra at \lm\ between bin 1 and all other $N_z$ bins.}\end{minipage}},\overbrace{ C^{\rm g}_{\ell_{\rm min}}(z_2,z_2),\ldots, C^{\rm g}_{\ell_{\rm min}}(z_2,z_{N_z})}^{\begin{minipage}{0.2\textwidth}\small{Auto- and cross-bin spectra at \lm\ between bin 2 and all other $N_z-1$ bins.}\end{minipage}}, C^{\rm g}_{\ell_{\rm min}+1}(z_1,z_1),\ldots,\underbrace{ C^{\rm g}_{\ell_{\rm max}}(z_{N_z},z_{N_z})}_{\begin{minipage}{0.13\textwidth}\small{Last of all the $N_\ell N_z(N_z+1)/2$ data points.}\end{minipage}}\bigg\},
\label{eq:data_vec}
\end{equation}
and then build the Gaussian log-likelihood as
\begin{equation}
-2\ln{L} =\sum_{\ell,\ell^\prime=\ell_{\rm min}}^{\ell_{\rm max}} \left\{\ln\left[2\upi\det\left(\boldsymbol\Gamma_{\ell\ell^\prime}\right)\right]+\left[\bm d_\ell-\bm t_\ell(\btheta)\right]^{\sf T}\left(\boldsymbol\Gamma_{\ell\ell^\prime}\right)^{-1}\left[\bm d_\ell-\bm t_\ell(\btheta)\right]\right\}.
\label{eq:likelihood}
\end{equation}
Here, $\bm t_\ell(\btheta)$ is the vector of the theoretical prediction based on a cosmological model defined by its cosmological parameters, whose values are stored in the parameter vector $\btheta$; the superscripts `${\sf T}$' and `$-1$' denote matrix transposition and inversion, respectively. This likelihood function is maximised for a given combination of values of the model parameters. In the current analysis, the Gaussian covariance matrix of \autoref{eq:covmat} is assumed to be independent on the parameters, and therefore the normilisation term of \autoref{eq:likelihood} can be ignored.

\subsection{Binning strategy}
\label{sec:CovFish}
To optimise our method, we adopt two binning strategies. First, we consider bins of the same size in redshift space (hereafter, `equi-spaced' bins), and then the case of bins with an equal number of galaxies in each (`equi-populated' bins). To choose among the two binning strategies presented in the previous section, i.e.\ equi-spaced vs equi-populated bins, we perform a preliminary Fisher matrix analysis \citep{Tegmark:1996bz}. Assuming a Gaussian likelihood for the cosmological parameters of interest, we can define the Fisher matrix $\mathbfss F$ with entries
\begin{equation}
F_{\alpha\beta}=\sum_{\ell,\ell^\prime=\ell_{\rm min}}^{\ell_{\rm max}}\frac{\partial  C^{\rm g}_\ell(z_i,z_j)}{\partial\theta_\alpha}\left(\boldsymbol\Gamma_{\ell\ell^\prime}^{-1}\right)^{ij,kl}\frac{\partial C^{\rm g}_{\ell^\prime}(z_k,z_l)}{\partial\theta_\beta},
\label{eq:Fisher}
\end{equation}
where $\theta_\alpha$ are the elements of the parameter vector $\btheta=\{\om,h,\sigma_8\}$.

\begin{figure*}
\centering
\includegraphics[width=0.45\textwidth]{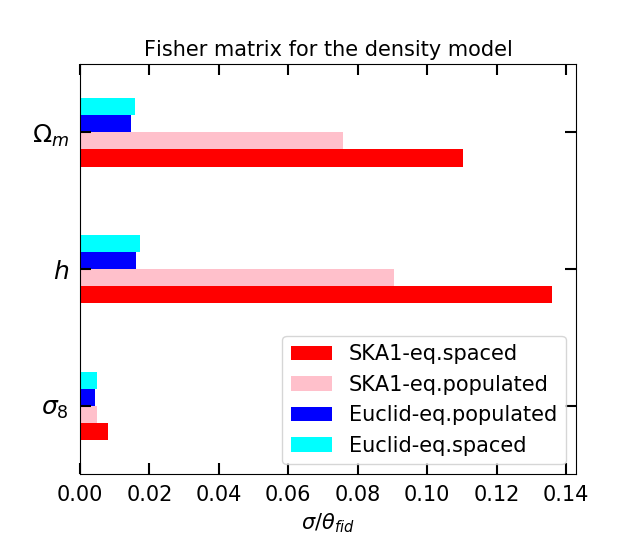}\includegraphics[width=0.45\textwidth]{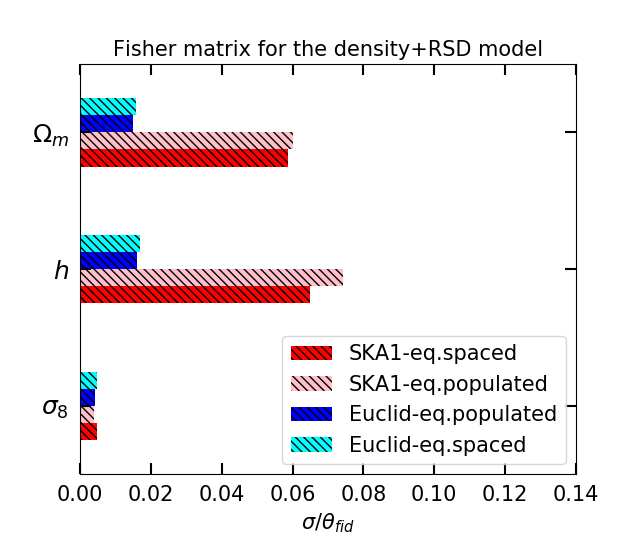}
\caption{Marginal $1\sigma$ Fisher matrix errors divided by the fiducial parameter value, for the two binning scenarios and the two proxy surveys. \textit{Left panel:} Model with density fluctuations only. \textit{Right panel:} Model with density fluctuations and RSD.}
\label{fig:Fisher_errors}
\end{figure*}

We forecast constraints on cosmological parameters by computing the Fisher matrix (in the appropriate multipole range) for both binning strategies, as well as for both $C^{\rm g,den}_{\ell\gg1}$ and $C^{\rm g,den+RSD}_{\ell\gg1}$. (Note that the covariance matrix in \autoref{eq:Fisher} is always the correct one, i.e.\ it includes both density fluctuations and RSD.)  Then, we compare the results. In \autoref{fig:Fisher_errors} we show the relative marginal errors on $\{\om,h,\sigma_8\}$ for all the cases considered. Constraints for \euc\ are always marginally tighter for equi-populated bins. In the case of SKA1, however, both binning strategies give almost equivalent results for the `den+RSD' model, whilst equi-populated bins yield tighter constraints for the `den' case. 

Overall, it is evident that the \euc-like survey is more constraining compared to SKA1. In order to investigate this, we calculate the cumulative signal-to-noise ratio (SNR) for the input reference cosmology,
\begin{equation}
    {\rm SNR}=\sqrt{\sum_{\ell,\ell^\prime=\ell_{\rm min}}^{\ell_{\rm max}} C^{\rm g}_\ell(z_i,z_j)\left(\boldsymbol\Gamma_{\ell\ell^\prime}^{-1}\right)^{ij,kl} C^{\rm g}_{\ell^\prime}(z_k,z_l)}.\label{eq:StoNRc}
\end{equation} 

\begin{figure*}
\centering
\includegraphics[width=0.77\textwidth]{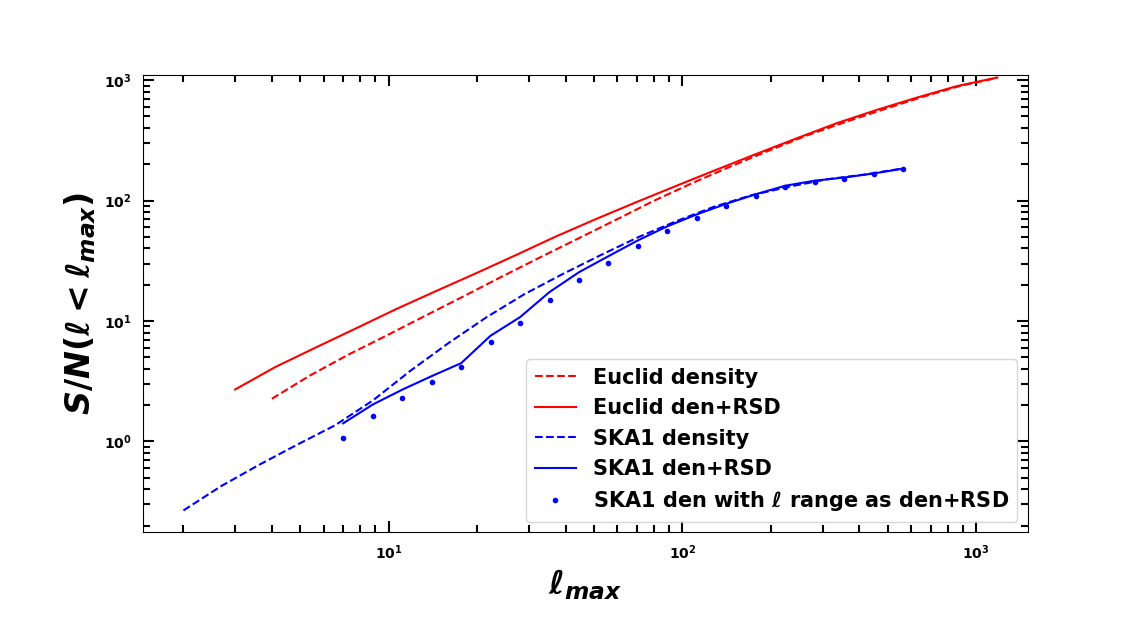}
\caption{Cumulative SNR as a function of the maximum multipole included in the analysis, \lM, for \euc\ and SKA1 equi-populated bins (red and blue curves, respectively) and the two models considered, i.e.\ den-only (dashed lines) and den+RSD (solid lines). The blue, dotted curve refers to the SKA1 SNR for density perturbations only in the case where we compute it in the same multipole range as for den+RSD.}
\label{fig:StoNRc}
\end{figure*}

 In \autoref{fig:StoNRc}, we present the cumulative SNR for \euc\ (red) and SKA1 (blue) with the `den-only' and `den+RSD' models (dashed and solid lines respectively) in the equi-populated scenario (this applies to the equi-spaced case as well). If we ignore for a while the different cumulative SNR between these two models within the same experiment, it is clear that generally the SNR for \euc\ is always greater than that of SKA1. The reason for this, is that \euc\, as seen in \autoref{tab:multipoles} extends to higher \lM values and also the sky fraction $f_{\rm sky}$ covered by this survey is three times that of SKA1. These two factors minimize the covariance matrix (see again \autoref{eq:likelihood}), yielding to an overall higher SNR. The specific features seen in \autoref{fig:StoNRc} will be discussed in more detail in \autoref{sec:ideal}.     

Furthermore, we perform preliminary MCMC tests for both surveys to make clear which binning configuration is computationally cheaper in terms of a faster convergence of the chains. Considering the case of density fluctuations and equi-populated bins, the chains converge quicker compared to equi-spaced bins for all the cases considered, whilst the convergence speed for den+RSD is comparable. Consequently, we conclude that the equi-populated redshift bins are more suitable to be adopted in the extensive and computationally expensive analysis of \autoref{sec:res}.

\begin{table}
\centering
\caption{Fiducial values and priors of cosmological and nuisance parameters.}
\begin{tabularx}{\textwidth}{Xllll}
	\hline
	 Parameter description & Parameter symbol & Fiducial value & Prior type & Prior range \\
	\hline
	\hline
	Present-day fractional matter density & $\om$ & 0.3089 & Flat & $[0.1,0.4]$\\
	Dimensionless Hubble parameter & $h$ & 0.6774 & Flat & $[0.5,1.0]$\\
    Amplitude of clustering$^\ddag$ & $\sigma_8$ & 0.8159 & Flat & $[0.5,1.2]$\\
	\hline
    Present-day fractional baryon density & $\ob$ & 0.0486 & -- & --\\
    Slope of the primordial curvature power spectrum & $\ns$ & 0.9667 & -- & --\\
	Amplitude of the primordial curvature power spectrum$^\ddag$ & $\ln(10^{10}\As)$ & $3.064$ & -- & --\\
    Optical depth to reionisation & $\tau_{\rm re}$ & 0.066 & -- & --\\
	\hline
    Photo-$z$ survey bias amplitude parameter$^\P$ & $\alpha_{\rm Euc}$ & 1.0 & Flat & $[0.6,1.4]$ \\
    Photo-$z$ survey bias slope parameter$^\P$ & $\beta_{\rm Euc}$ & 0.5 & Flat & $[0.1,0.9]$ \\
    Spectro-$z$ survey bias amplitude parameter$^\P$ & $\alpha_{\rm SKA}$ & 0.625 & Flat & $[0.2,1.0]$ \\
    Spectro-$z$ survey bias slope parameter$^\P$ & $\beta_{\rm SKA}$ & 0.881 & Flat & $[0.5,1.3]$ \\
	\hline
    Bin-dependent bias amplitude parameters$^\S$ & $b_{{\rm g},i}$ & 1.0 & Flat & $[0.1,1.9]$ \\
	\hline
\end{tabularx}\label{tab:params}
\raggedright\footnotesize{$^\ddag$ Adopting the LSS convention, we use $\sigma_8$ to parameterise the amplitude of matter fluctuations, thus setting the prior on this parameter rather than on the primordial amplitude parameter, $\As$.\\
$^\P$ Parameter varied in the reported prior range only in the `realistic scenario' of \autoref{sec:realistic}.\\
$^\S$ A dummy amplitude parameter for each redshift bin of \euc\ or SKA1 varied in the reported prior range only in the `conservative scenario' of \autoref{sec:conservative}.}
\end{table}

\section{Results and discussion}
\label{sec:res}
Throughout our analysis, in order to constrain the parameters of interest, we applied the Bayesian-based \emcee\ sampler \citep{ForemanMackey2013} and \Multinest\ \citep{FHB2009} interchangeably, depending on which sampling method is optimal/faster for each case. As discussed above, we focus on the set of cosmological parameters $\btheta=\{\om,h,\sigma_8\}$. Moreover, we also include a certain number of nuisance parameters, as described in the following three scenarios:
\begin{enumerate}
\item[$i)$] An ideal case where we constrain the cosmological parameter set assuming perfect knowledge of the galaxy bias;
\item[$ii)$] A realistic case with two bias nuisance parameters per experiment (see Eqs~\ref{eq:eucb} and \ref{eq:skaa});
\item[$iii)$] A conservative case where we include a nuisance parameter per redshift bin, thus allowing for a free redshift evolution of the bias.
\end{enumerate}
Reality is believed to lie between the last two cases. We note again that the procedure we follow is based on the rationale explained in \autoref{sec:CovFish}. That is, to create a mock data set where both density fluctuations and RSD are present, and then fit it against either a (wrong) model that ignores RSD, or a (correct) model that includes both density and RSD.

In an analysis where the \emcee\ or the \Multinest\ sampler is used, both high and the low likelihood areas are sampled, in contrast to the Fisher matrix, which only characterises the likelihood near its peak, assuming it is well approximated by a Gaussian. With our pipeline we want to explore the multi-dimensional parameter space of the two aforementioned models given the mock data in a Bayesian way. A major point in our analysis is the fact that we construct the mock data and, therefore, have perfect knowledge of the information it encodes. Hence, when we fit the mock data with the correct model, containing exactly the same information as the mock data, we expect this model to fit the data better than the wrong model, where the effect of the RSD in galaxy clustering is neglected. This latter, wrong model may or may not be sufficient to describe the data, depending mostly on the relative importance of signal, cosmic variance, and noise. In case it is proven not to be sufficient, the results will be biased. This bias will manifest as a misplaced peak in the posterior distribution. (Alternatively, it might also happen that the posterior exhibits some degree of bimodality.) In order to avoid referring to best-fit values---which can sometimes be misleading for strongly non-Gaussian posterior distributions---we opt for the means. The results of the pipeline analysis with \euc\ and SKA1 for the three scenarios discussed above are presented in \autoref{fig:main_ideal},  \autoref{fig:main_realistic}  and \autoref{fig:main_conservative}, respectively. \autoref{tab:results_Euclid} and \autoref{tab:results_SKA1} list estimates of the means and $68\%$ marginal errors on each parameter. We discuss these results thoroughly in the following subsections.

\subsection{Ideal scenario}\label{sec:ideal}
\begin{figure*}
\centering
\includegraphics[width=0.33\textwidth]{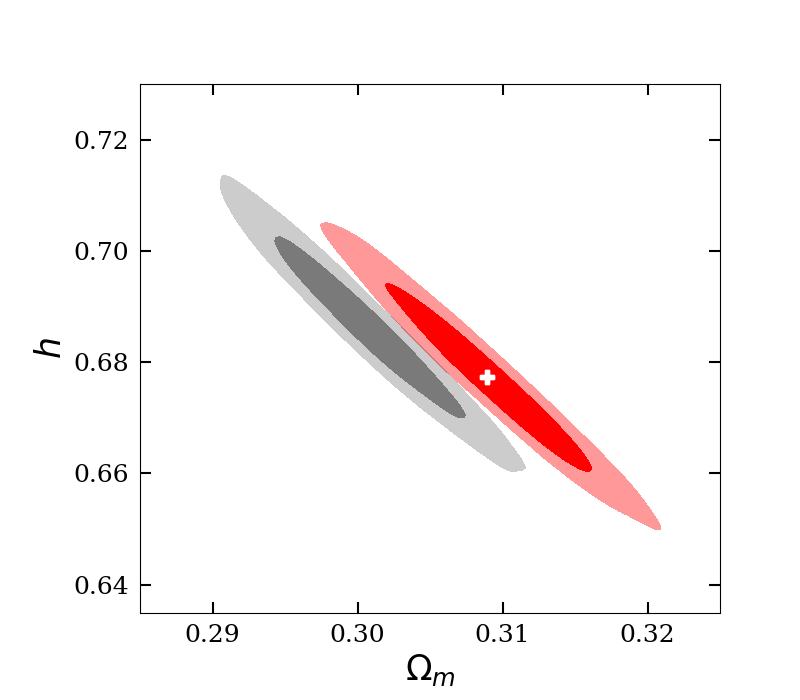}\includegraphics[width=0.33\textwidth]{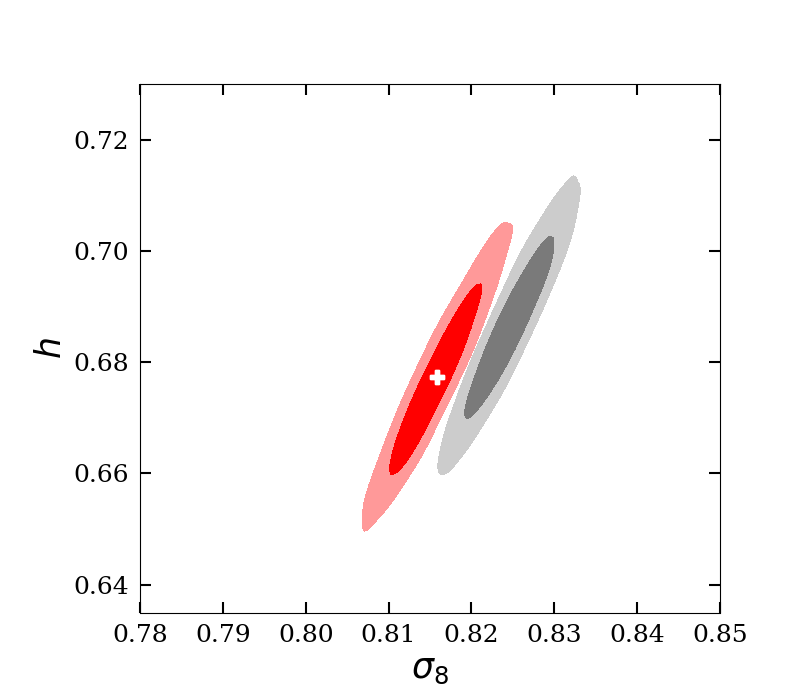}\includegraphics[width=0.33\textwidth]{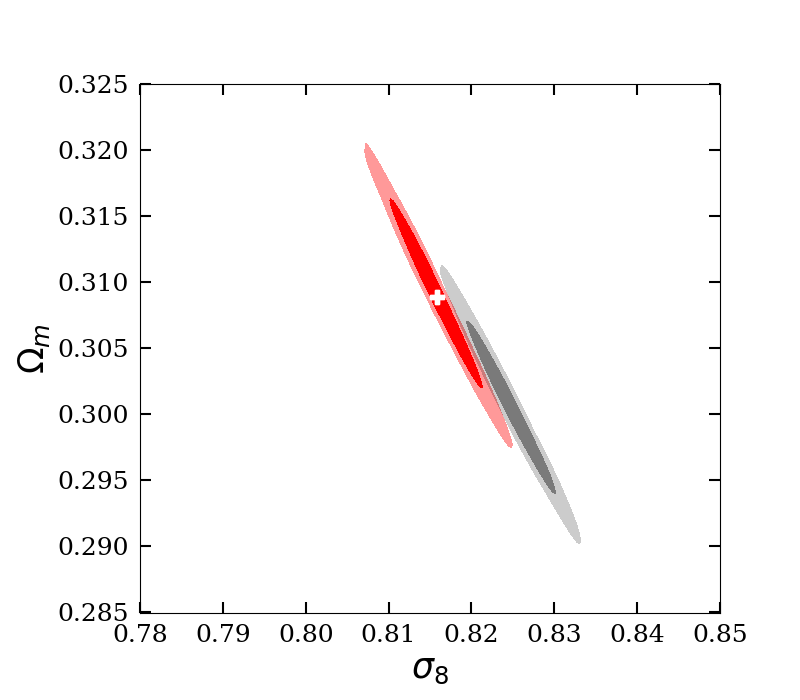}\\
\includegraphics[width=0.33\textwidth]{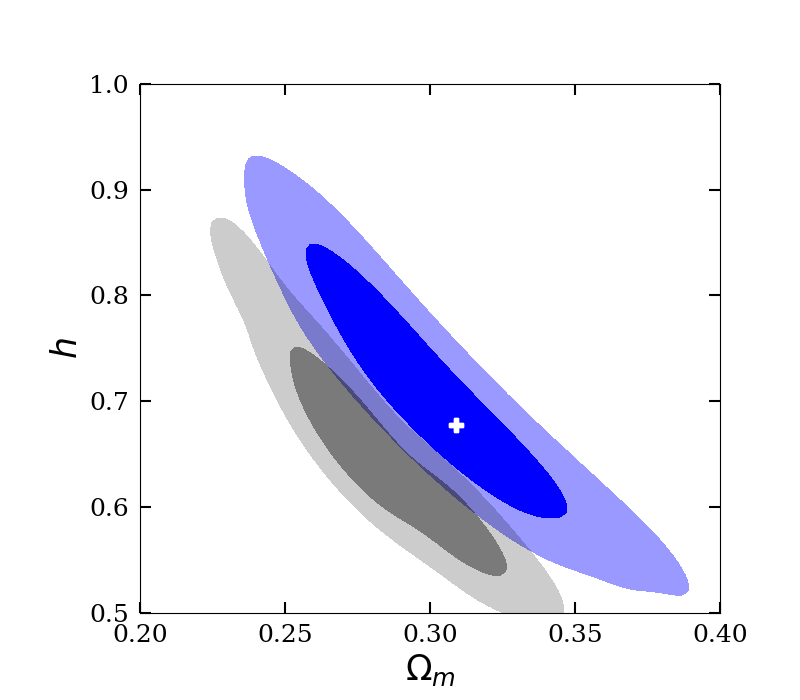}\includegraphics[width=0.33\textwidth]{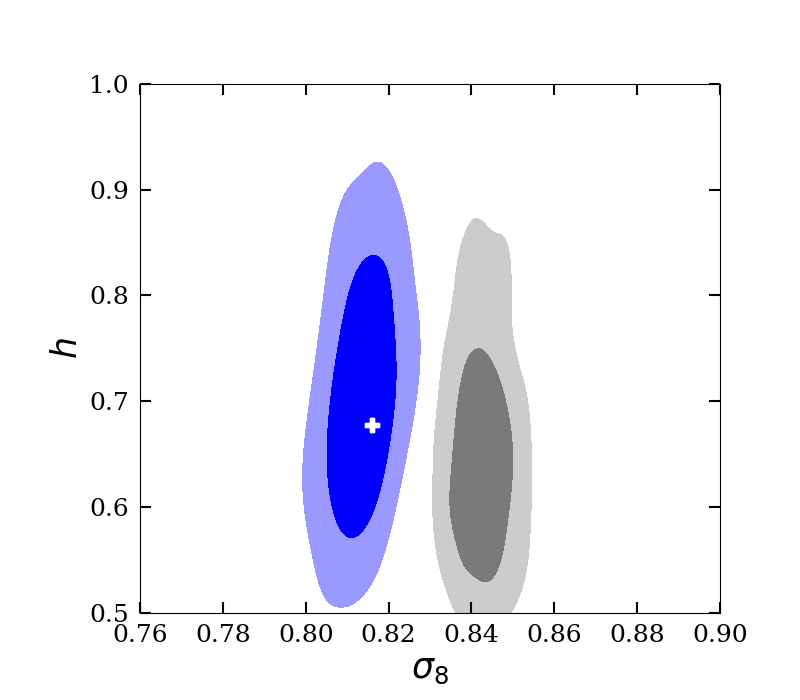}\includegraphics[width=0.33\textwidth]{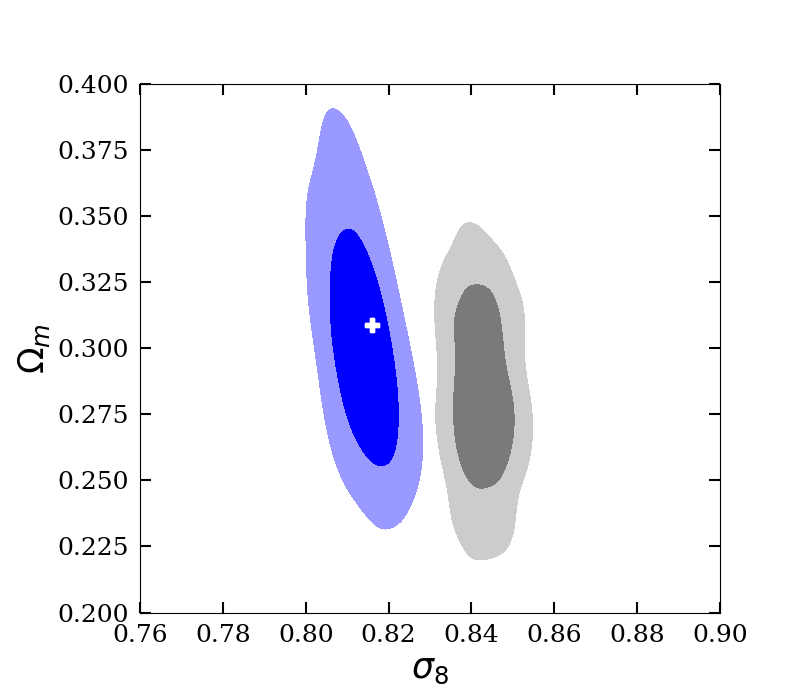}
\caption{Constraints on cosmological parameters for the ideal case, i.e.\ no nuisance parameters. Outer and inner contours respectively correspond to 95\% and 68\% confidence levels in the joint 2D parameter space. \textit{Top panels:} (\textit{Bottom panels:}) parameter estimation from the \euc-like optical/near-infrared photometric (SKA1-like radio spectroscopic \hi-line) galaxy survey with the red (blue) and grey contours accounting for the complete and the incomplete model respectively. The white cross indicates the fiducial cosmology.}
\label{fig:main_ideal}
\end{figure*}

\begin{figure*}
\centering
\includegraphics[width=0.36\textwidth]{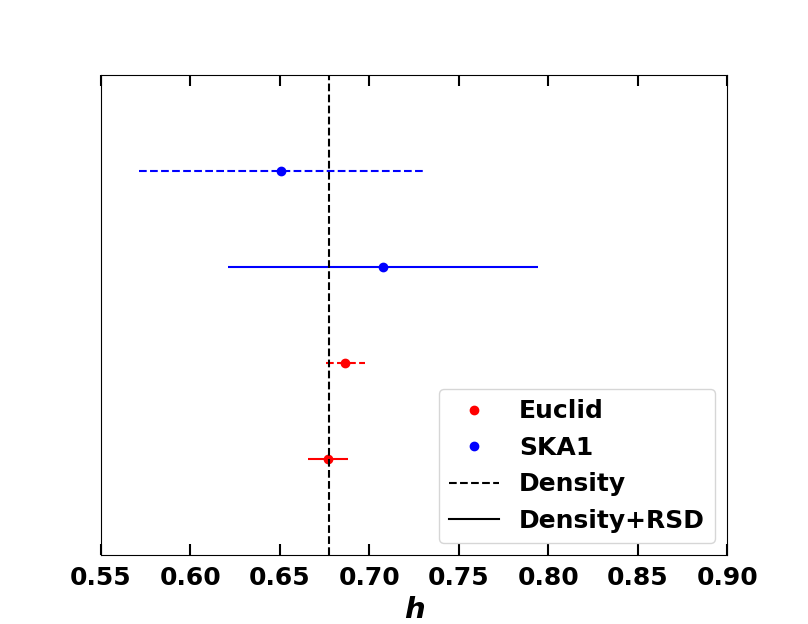}\includegraphics[width=0.36\textwidth]{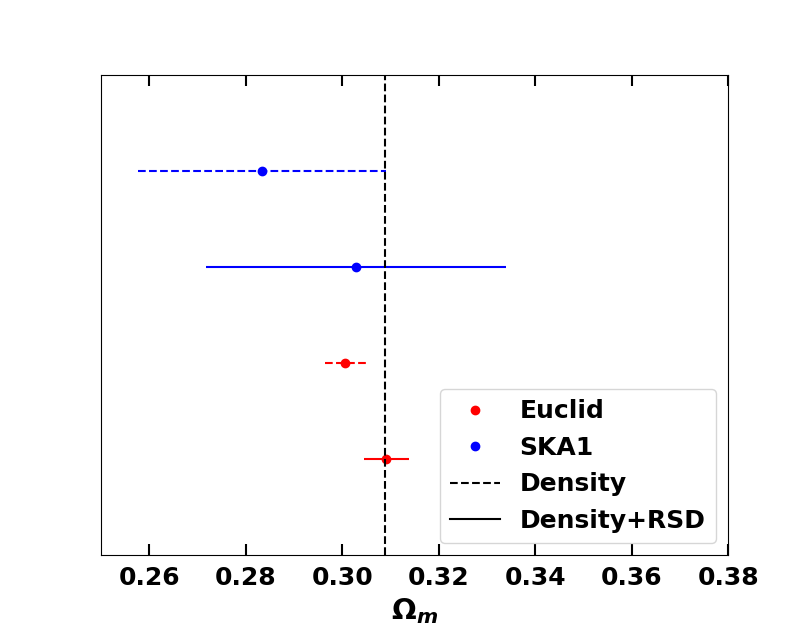}\includegraphics[width=0.36\textwidth]{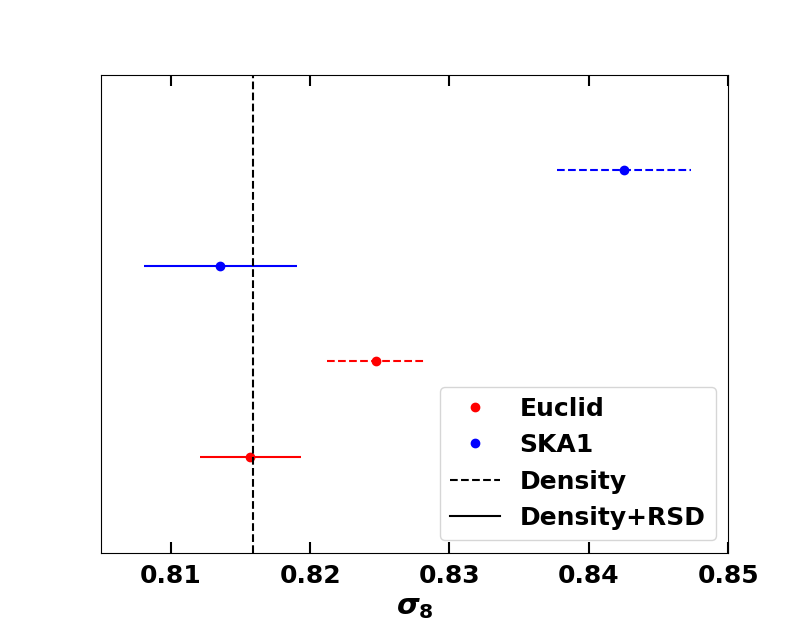}
\caption{$1\sigma$ marginal errors (horizontal solid (complete model) and dashed (incomplete model) lines) on the estimated mean value (filled bullets) for \euc\ and SKA1 in the ideal scenario. The vertical dashed black line corresponds to the input value of our fiducial cosmology. }
\label{fig:errorsideal}
\end{figure*}

In \autoref{fig:main_ideal} (top panels) we show the $68\%$ and $95\%$ joint marginal error contours for the Bayesian analysis with \euc\ on the parameter set $\{\om,h,\sigma_8\}$. We use priors and fiducial values as given in \autoref{tab:params}. These constraints appear quite stringent, and it is clear that, when we fit the mock data with the correct model (in red), the input reference cosmology (white cross) lies well within the $1\sigma$ regions of the reconstructed parameter error intervals. On the contrary, if we assume the wrong data model---namely we do not include RSD in the theoretical data vector---it is evident that the reconstructed contours (in grey) are biased with respect to the input cosmology. It is worth noticing that the 2$\sigma$ regions do not overlap in parameter space. This may seem somewhat unexpected, as it is often assumed that RSD do not matter when one deals with photometric galaxy surveys. However, this finding, which represents one of the main results of our paper, is also in agreement with previous literature focussed on galaxy clustering including RSD for photometric redshifts \citep[e.g.][]{Padmanabhan2006,Blake2006,Crocce2010}. For instance, \citet{Nock2010} proposed a new binning scheme based on galaxy pair centres rather than the galaxy positions, to alleviate the anisotropic RSD on the projected galaxy two-point function.  This is more evident in \autoref{fig:errorsideal}, where the estimated mean for the incomplete model (red bullet point) is more than $1\sigma$ away (red, dashed line) from the input values of parameters $\{\om,\sigma_8\}$, shown as vertical dashed black lines. 

Similarly, in \autoref{fig:main_ideal} (bottom panels) we present the constraints on the parameters from the SKA1. In particular, SKA1 yields weaker constraints than \euc\ due to the  lower SNR (see \autoref{fig:StoNRc}) , as discussed at the end of \autoref{sec:CovFish}, namely the smaller $f_{\rm sky}$ and the more limited multipole range. In this case, too, it is evident that the estimate from the incomplete, density-only model is biased beyond $1\sigma$ for all cosmological parameters, whereas results from the den+RSD model are consistent with the input cosmology (see again \autoref{fig:errorsideal} the blue lines). However, we find that den+RSD model yields slightly weaker constraints compared to the (biased) ones we get when neglecting RSD.

In order to understand this we need to go back to \autoref{fig:StoNRc}.
In this plot as previously seen in \autoref{sec:CovFish} the SNR is shown, with red and blue curves respectively referring to \euc\ and SKA1, and dashed(solid) lines for den-only(den+RSD); we also show, as a blue dotted curve, the SKA1 cumulative SNR for den-only in the case where we use the same multipole range as for den+RSD. We notice that for \euc\ the SNR curve corresponding to den+RSD is always higher than that of the density fluctuations only in the whole multipole range. This makes sense, since we consider additional information by adding the RSD on top of the density fluctuations and, as a result, we increase the signal and obtain higher SNR. Regarding the SKA1 setup, the SNR curves will be significantly lower than in the case of \euc\  for the reasons explained in \autoref{sec:CovFish}. By looking the SNR, we see that the curve for the complete (density+RSD) model is below that of the incomplete one, which neglects RSD. This trend seems to be the exact opposite of the what discussed for \euc. However, we should note that in the case of SKA1 the multipole range where we can trust the Limber approximation is smaller for density+RSD, compared to density perturbations only (see \autoref{tab:multipoles}). Given that, we compute again the SNR of the density model but now evaluated at the shorter multipole range that was applied for the correct model. After implementing this (dotted curve), we now observe the same trend as for \euc. This implies that the relatively larger contours for SKA1 den+RSD have to be attributed to the higher $\ell_{min}$ limit resulting in a slightly shorter multiple range.

\subsection{Realistic scenario}\label{sec:realistic}
As mentioned at the beginning of the section, the assumption that our knowledge of the galaxy bias is perfect is an idealistic one. Thus, we now introduce nuisance parameters to account for our inherent ignorance of the bias. Such parameters will then be fitted alongside cosmological parameters. To this purpose, we choose a similar modelling for the two surveys under consideration, i.e.\ an overall normalisation of the galaxy bias over the whole redshift range, and a parameter accounting for the redshift dependence of the bias. In other words, we let the parameters $\alpha_X$ and $\beta_X$ of \autoref{eq:eucb} and \autoref{eq:skaa} to vary freely, with $X=\{{\rm Euc},{\rm SKA}\}$. The normalisation and power-law bias nuisance parameters with their corresponding priors for the surveys are shown in \autoref{tab:params}.

\begin{figure*}
\centering
\includegraphics[width=0.33\textwidth]{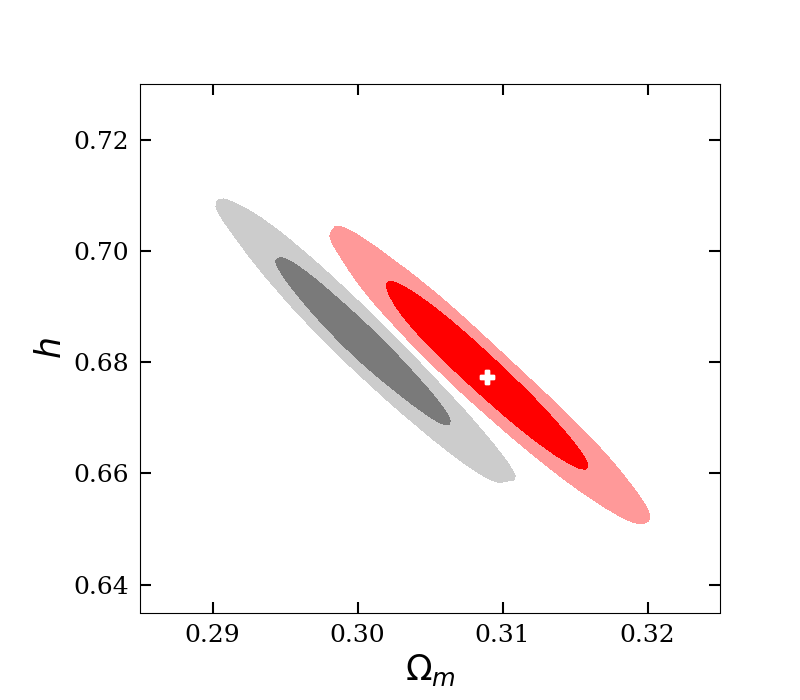}\includegraphics[width=0.33\textwidth]{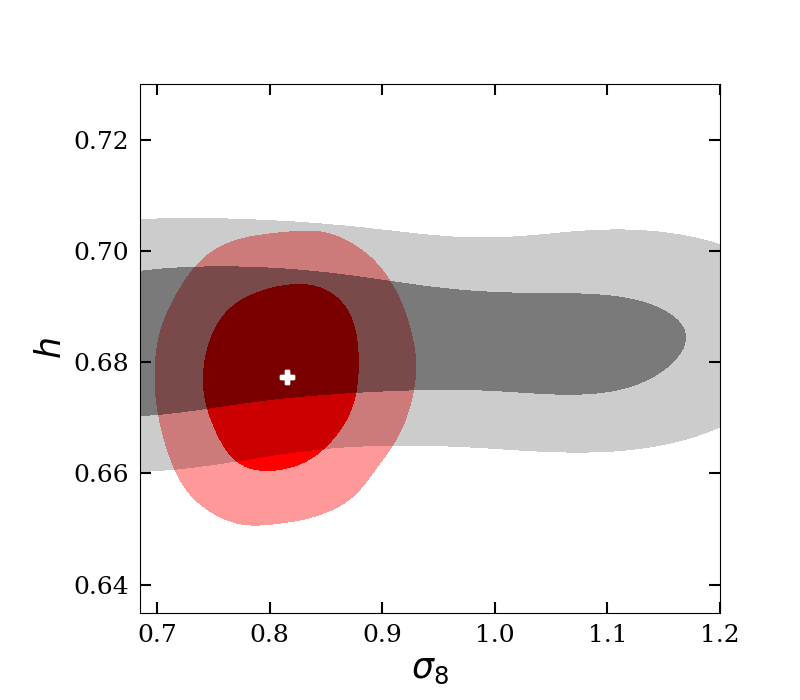}\includegraphics[width=0.33\textwidth]{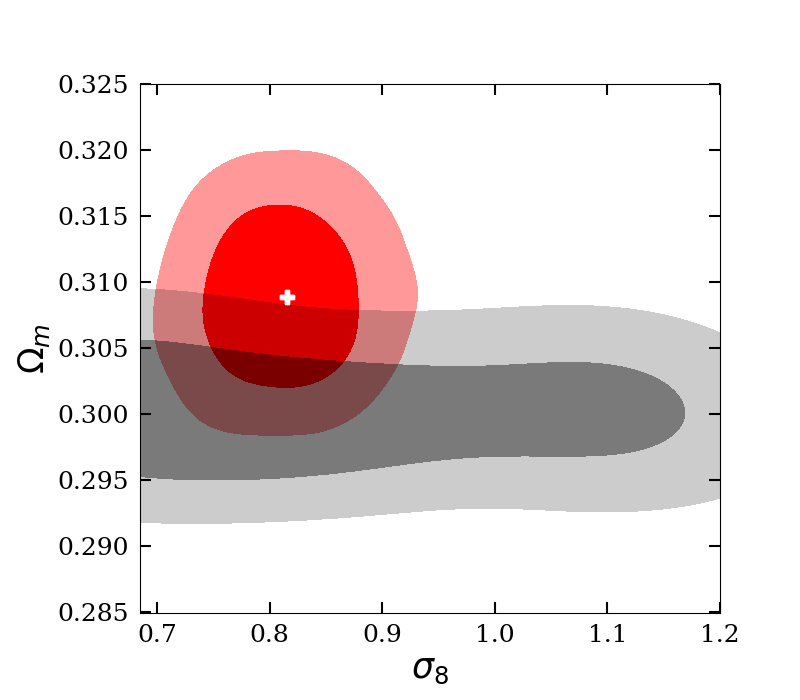}\\
\includegraphics[width=0.33\textwidth]{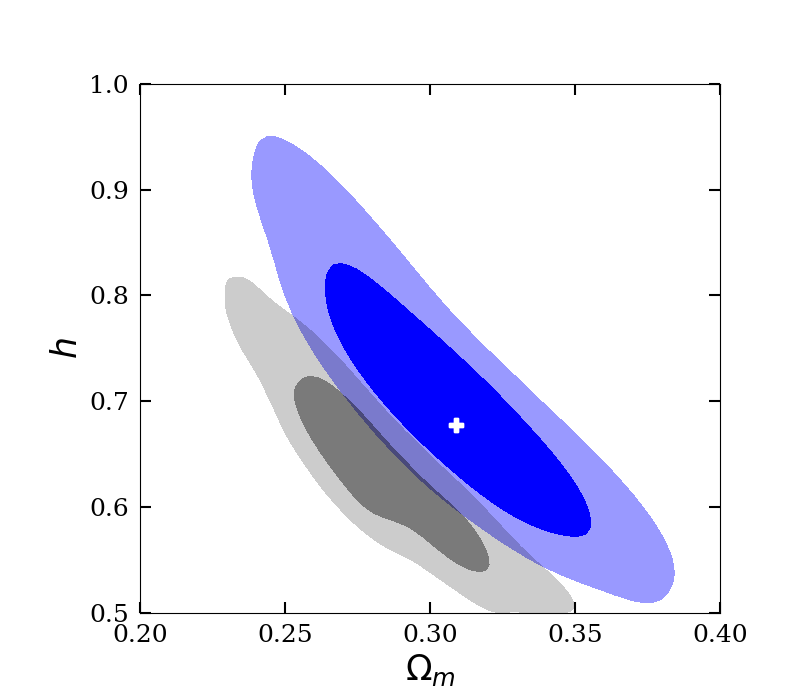}\includegraphics[width=0.33\textwidth]{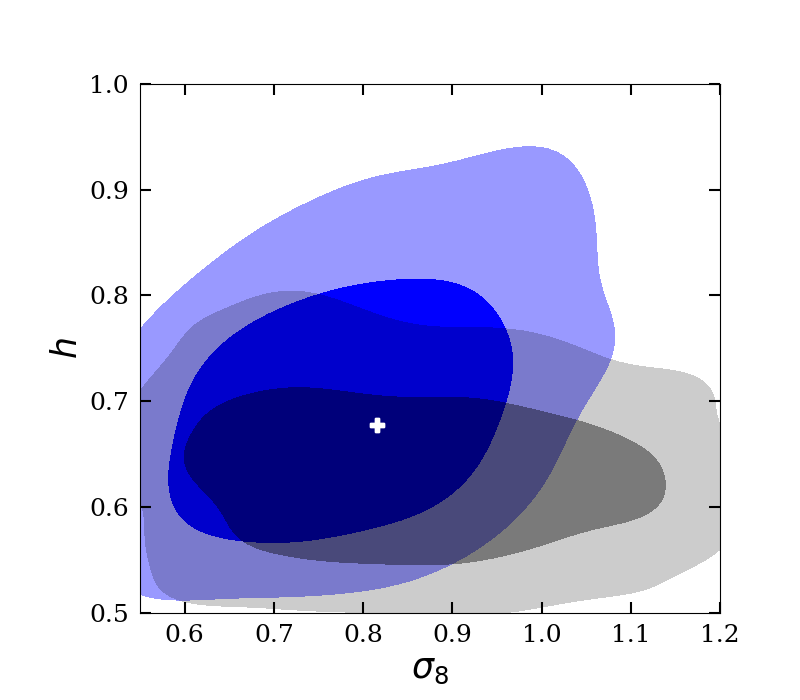}\includegraphics[width=0.33\textwidth]{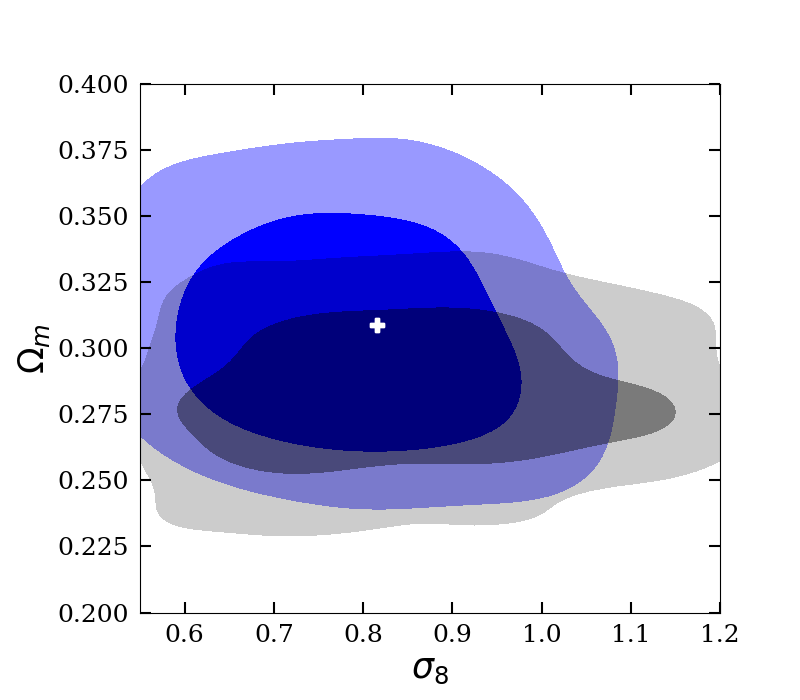}
\caption{Same as \autoref{fig:main_ideal} but for the realistic scenario, i.e.\ two nuisance parameters modelling the overall amplitude and the redshift evolution of the bias.}
\label{fig:main_realistic}
\end{figure*}

\autoref{fig:main_realistic} (top panels) shows the results for the optical/near-infrared \euc-like photometric survey, after marginalising over bias nuisance parameters. Interestingly, the constraints on $h$ and $\om$ are very similar to those of the ideal scenario. That is, the biased estimate for density only lies beyond $1\sigma$ on $\om$ but not for $h$ with respect to the fiducial values. However, the picture is completely different when it comes to $\sigma_8$. It is clear that $\sigma_8$ is totally unconstrained by the density-only model (grey contours). The reason for this is that density fluctuations are sensitive to the galaxy bias (the angular power spectrum depends linearly on the bias squared). This means that when we consider an overall normalisation of the bias---common to the whole redshift range---we cannot break the degeneracy present between $\alpha_X$ and $\sigma_8$. On the other hand, once we include RSD (red contours), the degeneracy is lifted almost completely.

The SKA1 results for this realistic bias scenario are shown in the bottom panels of \autoref{fig:main_realistic}. We can appreciate a similar behavior compared to the case of \euc. The incomplete model containing only density fluctuations is statistically significantly biased on $\om$ and, again, the constraint on $\sigma_8$ is very degenerate for the reasons explained above. By incorporating RSD in our modeling we manage to alleviate this and get an unbiased estimate of $\om$. Again, the constraining power of SKA1 is not so good as that of the \euc-like survey, due to the lower SNR.
\begin{figure*}
\centering
\includegraphics[width=0.36\textwidth]{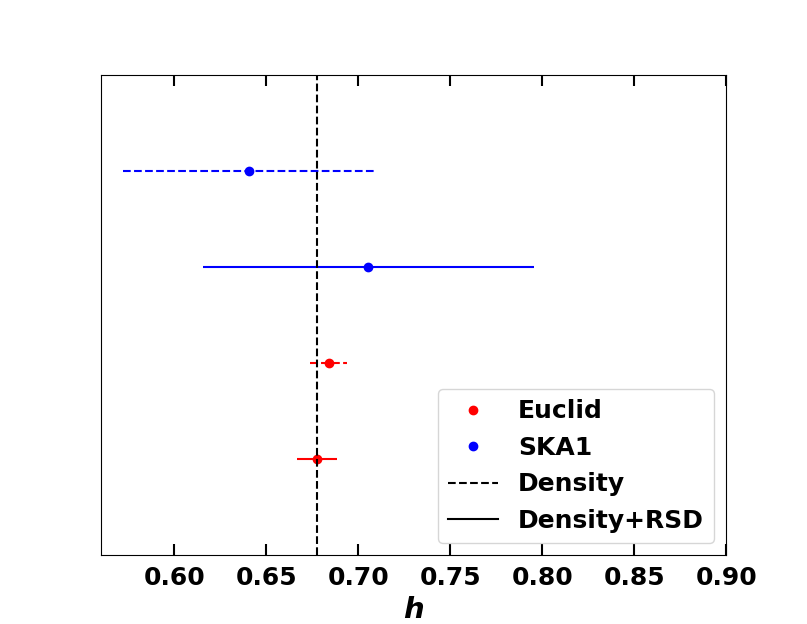}\includegraphics[width=0.36\textwidth]{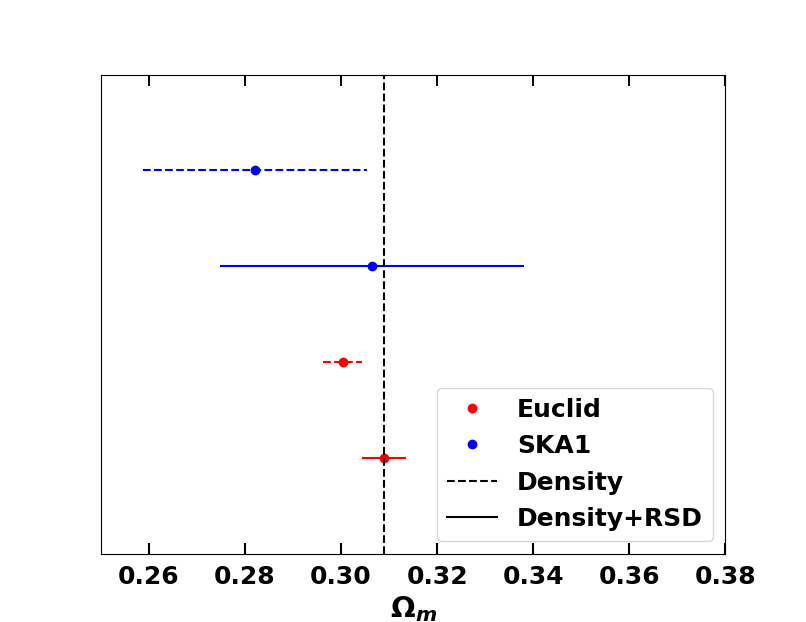}\includegraphics[width=0.36\textwidth]{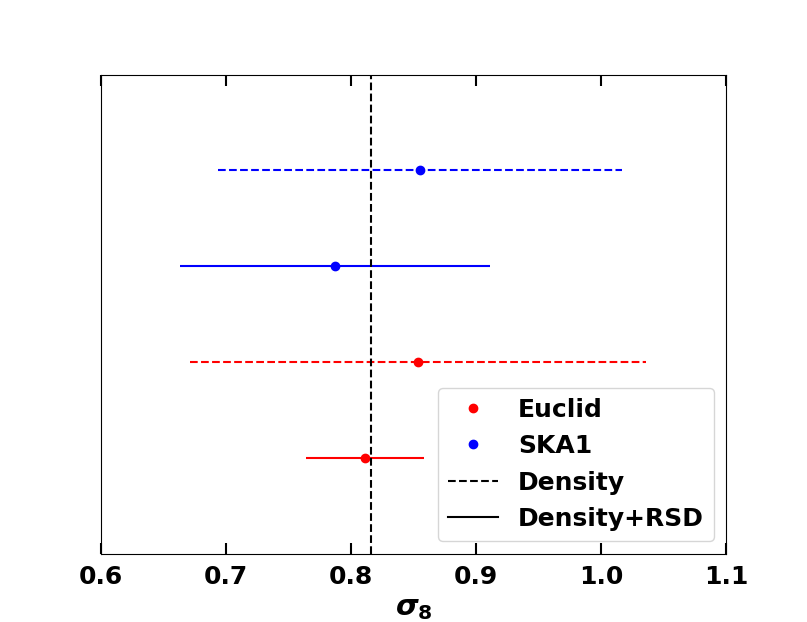}\\
\caption{Same as \autoref{fig:errorsideal} but for the realistic scenario}
\label{fig:errorsreal1}
\end{figure*}

\subsection{Conservative scenario}\label{sec:conservative}
Let us now consider the pessimistic case in which the galaxy bias evolution with redshift is utterly unknown. Thus, we add  bias nuisance parameters per redshift bin $b_{{\rm g},i}$, with $i=1,N_z$, and flat priors in the range $[0.1, 1.9]$. We then obtain constraints over the full parameter set consisting of 13 parameters---namely three cosmological parameters plus $N_z$ bias nuisance parameters---and present the joint 2D marginal error contours on the cosmological parameters by marginalising over all the bias parameters.
\begin{figure*}
\centering
\includegraphics[width=0.33\textwidth]{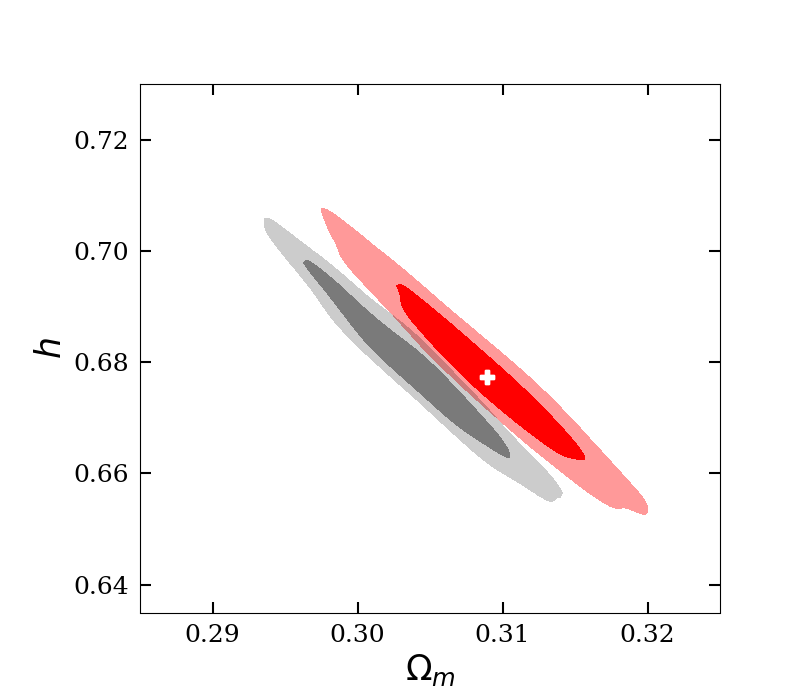}\includegraphics[width=0.33\textwidth]{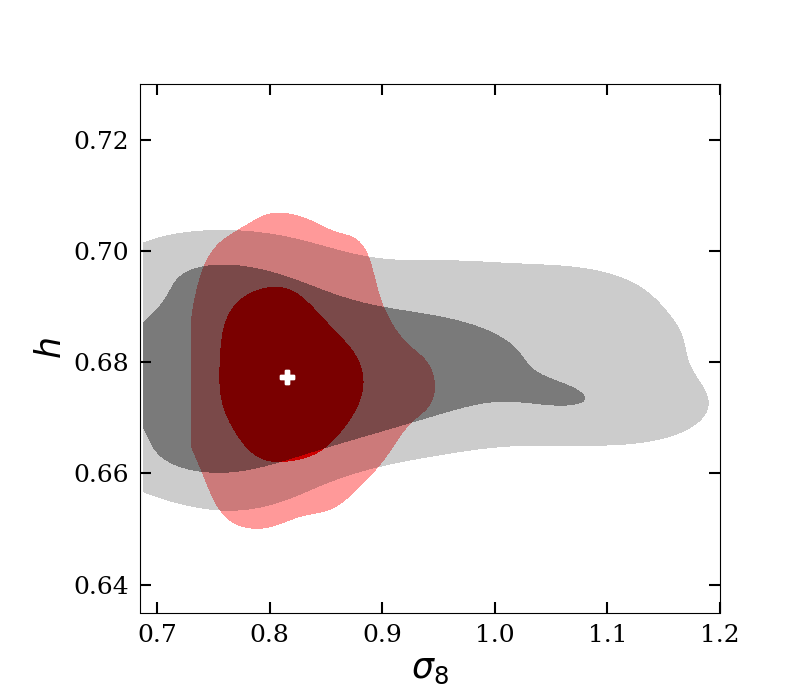}\includegraphics[width=0.33\textwidth]{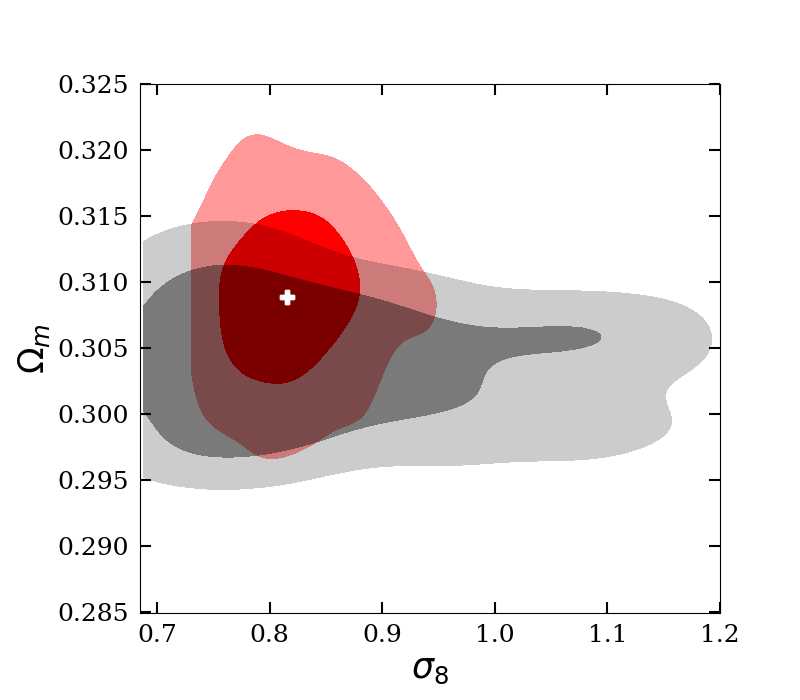}\\
\includegraphics[width=0.33\textwidth]{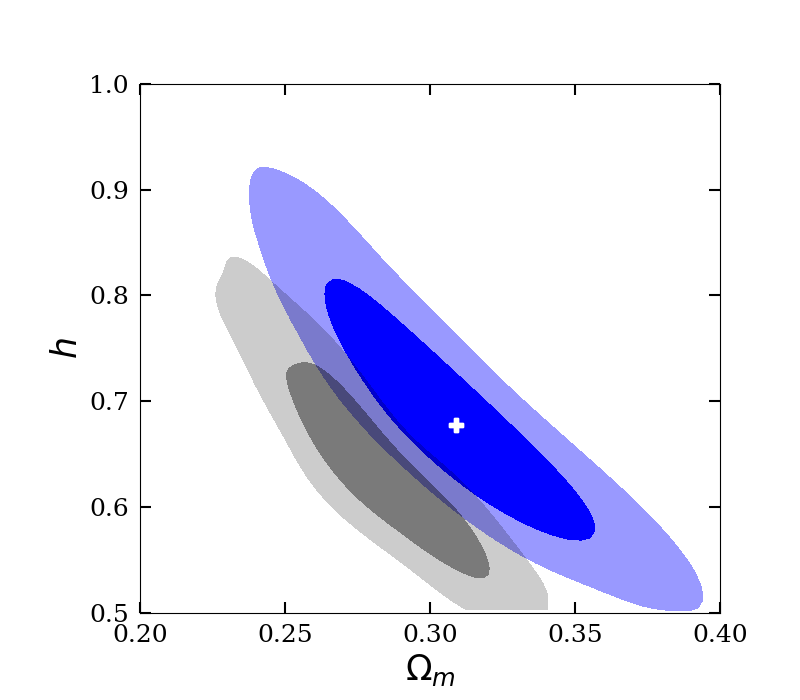}\includegraphics[width=0.33\textwidth]{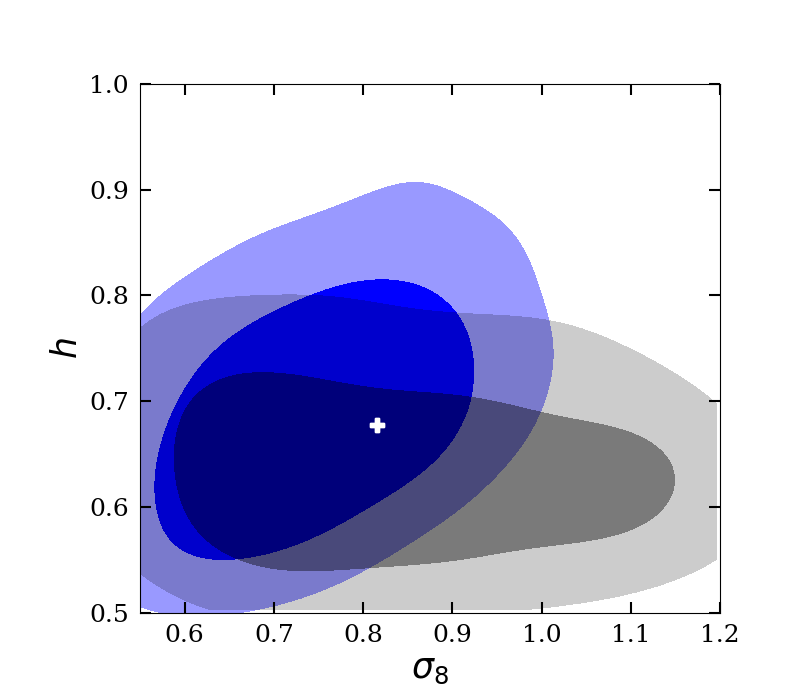}\includegraphics[width=0.33\textwidth]{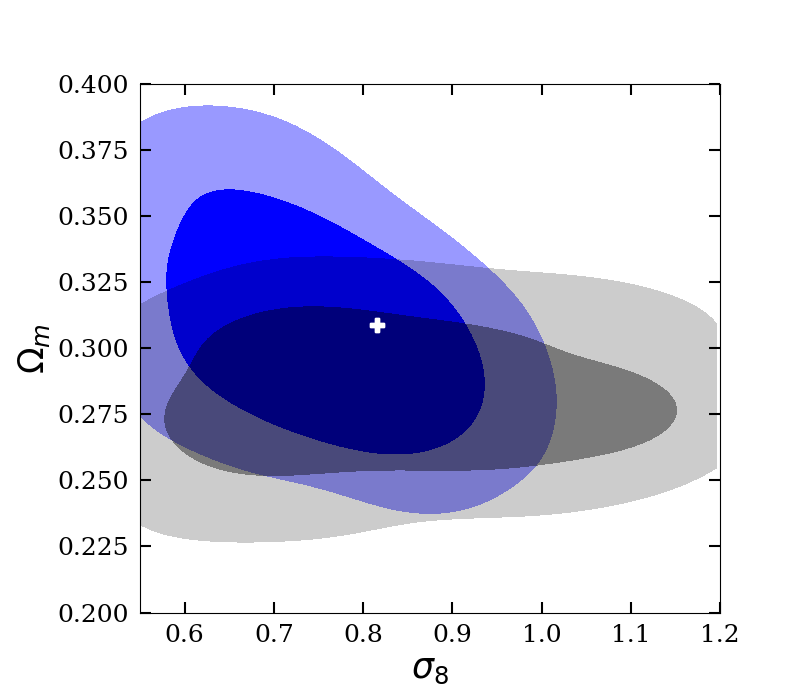}
\caption{Same as \autoref{fig:main_ideal} but for the conservative scenario, i.e.\ one nuisance bias parameter per redshift bin.}
\label{fig:main_conservative}
\end{figure*}

As before, in \autoref{fig:main_conservative} (top panels) we present the cosmological constraints from \euc. Again, we can clearly see that the results on $h$ and $\om$ are quite similar to those from the ideal and the realistic scenario with the matter density parameter $\om$ being more than $1\sigma$ away from the input values for the incomplete model. Likewise, the results on the normalisation $\sigma_8$ are equivalent to that of the pessimistic case. That is, the persistence of the degeneracy on $\sigma_8$. We, again, alleviate this with the correct den+RSD model---since RSD are not sensitive to the galaxy bias---which yields results in agreement with the fiducial cosmology.

The case for SKA1 is  shown in \autoref{fig:main_conservative} (bottom panels). It is obvious, as well, that the picture does not change with respect to the pessimistic scenario. In a similar fashion, the incomplete model yields degenerate results on $\sigma_8$, while the correct model gives more tighter constraints. In addition to that, the estimate of the density model on $\Omega_m$ remains biased more than $1\sigma$ away.  
\begin{table*}
\centering
\caption{Means and corresponding $68\%$ marginal error intervals on cosmological parameters for the optical/near-infrared \euc-like photo-$z$ galaxy survey.}
\begin{tabular}{lllcllcll}
    \hline
    \multicolumn{9}{c}{\euc} \\
    \hline
   & \multicolumn{2}{c}{Ideal scenario} && \multicolumn{2}{c}{Realistic scenario} && \multicolumn{2}{c}{Conservative scenario} \\
    \cline{2-3}\cline{5-6}\cline{8-9}
   & \multicolumn{1}{c}{den} & \multicolumn{1}{c}{den+RSD} && \multicolumn{1}{c}{den} & \multicolumn{1}{c}{den+RSD} && \multicolumn{1}{c}{den} & \multicolumn{1}{c}{den+RSD} \\
    \hline
    \hline
    $\om$ & $0.3006\pm0.0042$ & $0.3091\pm0.0046$ && $0.3003\pm0.0040$ & $0.3089\pm0.0045$ && $0.3038\pm0.0042$ & $0.3089\pm0.0045$ \\
    $h$ & $0.6865\pm0.0107$ & $0.6770\pm0.0111$ && $0.6837\pm0.0101$ & $0.6775\pm0.0109$ && $0.6791\pm0.0105$ & $0.6778\pm0.0108$ \\
    $\sigma_8$ & $0.8247\pm0.0034$ & $0.8157\pm0.0036$ && $0.8534\pm0.1823$ & $0.8111\pm0.0474$ && $0.859\pm0.1298$ & $0.8211\pm0.0469$ \\
    \hline
\end{tabular}
\label{tab:results_Euclid}
\end{table*}

\begin{table*}
\centering
\caption{Means and corresponding $68\%$ marginal error intervals on cosmological parameters for the radio SKA1-like spectro-$z$ galaxy survey.}
\begin{tabular}{lllcllcll}
    \hline
    \multicolumn{9}{c}{SKA1} \\
    \hline
   & \multicolumn{2}{c}{Ideal scenario} && \multicolumn{2}{c}{Realistic scenario} && \multicolumn{2}{c}{Conservative scenario} \\
    \cline{2-3}\cline{5-6}\cline{8-9}
   & \multicolumn{1}{c}{den} & \multicolumn{1}{c}{den+RSD} && \multicolumn{1}{c}{den} & \multicolumn{1}{c}{den+RSD} && \multicolumn{1}{c}{den} & \multicolumn{1}{c}{den+RSD} \\
    \hline
    \hline
    $\om$ & $0.2833\pm0.0256$ & $0.3028\pm0.0311$ && $0.2821\pm0.0232$ & $0.3063\pm0.0316$ && $0.2811\pm0.0239$ & $0.3084\pm0.0329$ \\
    $h$ & $0.6504\pm0.0793$ & $0.7077\pm0.0866$ && $0.6404\pm0.0684$ & $0.7054\pm0.0897$ && $0.6443\pm0.0752$ & $0.6887\pm0.0857$ \\
    $\sigma_8$ & $0.8425\pm0.0048$ & $0.8135\pm0.0055$ && $0.8552\pm0.1613$ & $0.7872\pm0.1238$ && $0.8438\pm0.1703$ & $0.7467\pm0.1118$ \\
    \hline
\end{tabular}
\label{tab:results_SKA1}
\end{table*}

\begin{figure*}
\centering
\includegraphics[width=0.36\textwidth]{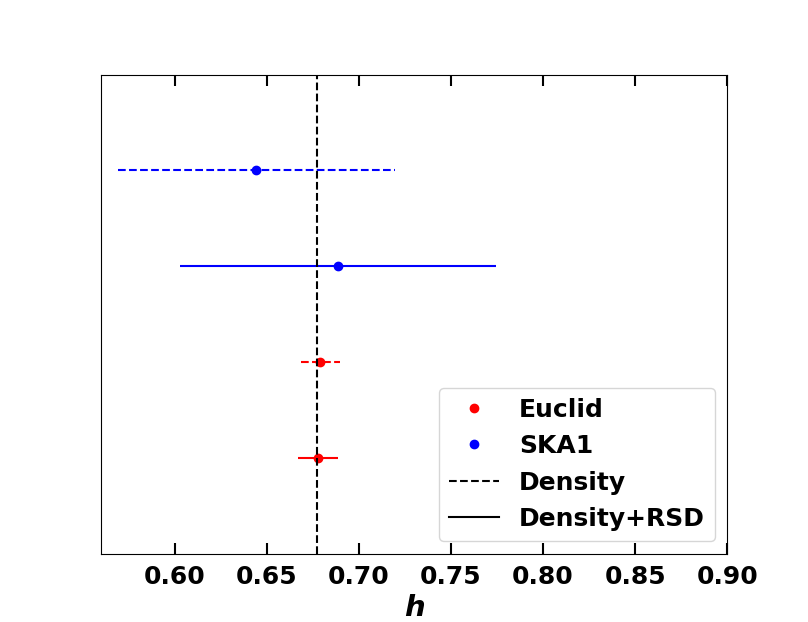}\includegraphics[width=0.36\textwidth]{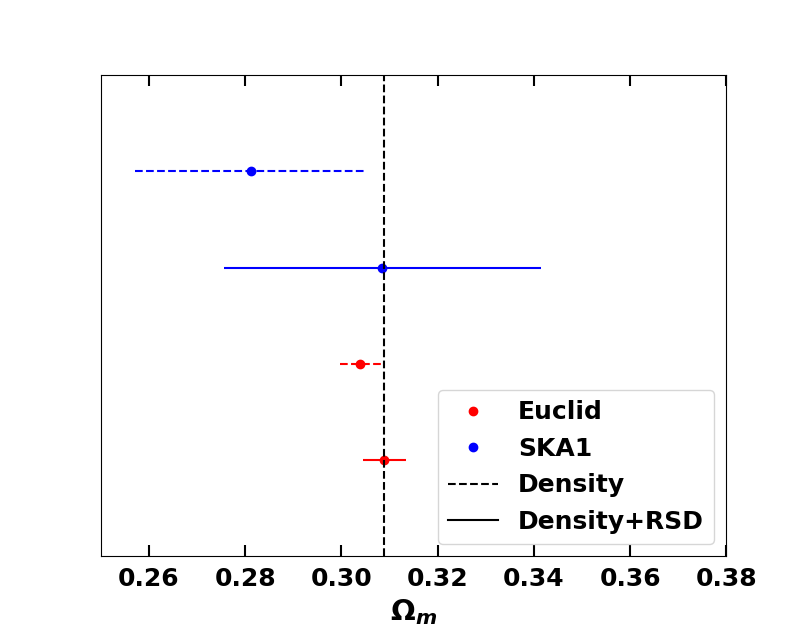}\includegraphics[width=0.36\textwidth]{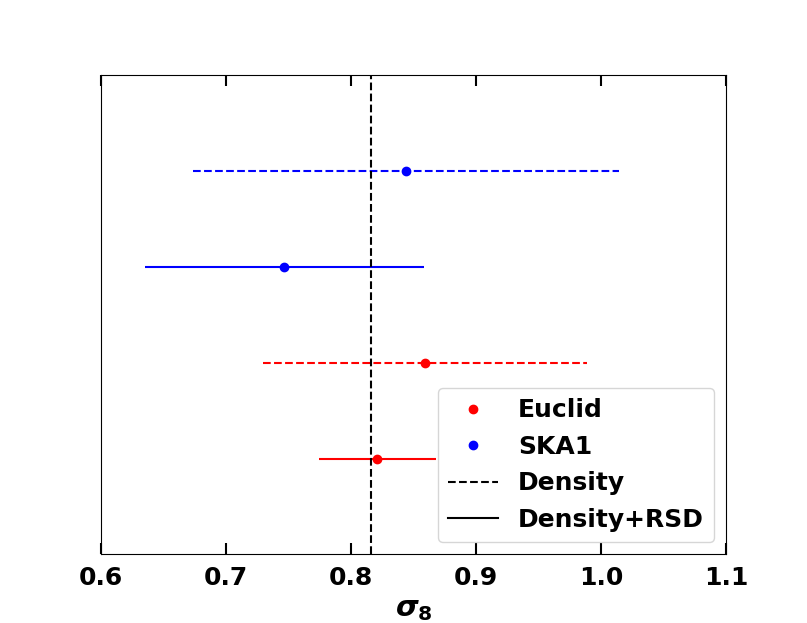}\\
\caption{Same as \autoref{fig:errorsideal} but for the conservative scenario}
\label{fig:errorsreal2}
\end{figure*}

\section{Conclusions}
\label{sec:disc}
In this work, we have studied the effect of redshift-space distortions (RSD) on the tomographic angular power spectrum of galaxy number count fluctuations (in the linear regime). In detail, we estimated to what extent the information encoded in the RSD term can affect a cosmological analysis. To this purpose, we have introduced, for the first time to our knowledge, the RSD along with the density perturbations in the Limber approximation. We have modified the publicly available \cosmosis\ code, and we have validated it at given redshift and multipole ranges against the Boltzmann solver code \class.

In order to study the impact of RSD, we have followed this rationale. First, we construct mock observables in the form of galaxy number count tomographic angular power spectra, \tcl{i}{j},  including both density fluctuations and RSD. Then, we fit this synthetic data with two theoretical models:
\begin{itemize}
    \item A model that incorporates exactly the same information as in the mock data set;
    \item A model that ignores RSD.
\end{itemize}

For this analysis, we have adopted two planned galaxy surveys, one as a proxy for future photometric missions in the optical/near-infrared waveband, and another as a representative of oncoming spectroscopic experiments at radio frequencies. The former follows the specifications of a \euc-like satellite, whereas for the latter we have considered \hi-line galaxy observations as performed by SKA1 (the first phase of the SKA radio telescope). In order to opt between an equi-populated and an equi-spaced redshift binning, we have performed a Fisher matrix test and a preliminary MCMC analysis on the cosmological set $\{\om,h,\sigma_8\}$. After choosing the former as the optimal binning configuration, we have proceeded to a more extensive Bayesian analysis. For the final analysis, we have considered:
\begin{enumerate}
    \item[$i)$] An ideal scenario, with no nuisance parameter to model the galaxy bias;
    \item[$ii)$] A realistic scenario, with an overall normalisation and a redshift dependence to account for a certain ignorance of the bias;
    \item[$iii)$] A conservative scenario, where the bias can evolve freely over the redshift range.
\end{enumerate}

Given these cases we can summarise our basic results as:
\begin{itemize}
    \item The discrepancy on the estimated mean values of cosmological parameters between an analysis with and without RSD is statistically significant for both our proxy surveys, especially for the parameters $\{\om,\sigma_8\}$. This holds true for both the ideal, the realistic and the conservative scenario (see \autoref{fig:errorsideal}, \autoref{fig:errorsreal1} and \autoref{fig:errorsreal2}).
    \item The wrong theoretical model (including only density perturbations) yields very degenerate results on $\sigma_8$, since the normalisation of the matter power spectrum and the overall normalisation of the bias are completely degenerate. This happens in a similar fashion when we consider bias nuisance parameters per redshift bin. We partially lift this degeneracy when we add RSD, which are insensitive to the galaxy bias. 
    \item Overall, SKA1 is less informative than \euc\ due to the lower SNR ascribed to the shorter multipole range and the smaller sky coverage.
\end{itemize}

These results demonstrate that the inclusion of RSD on top of the density fluctuations in our theoretical predictions is of great importance in order to avoid large biases which dominate the statistics and inevitably lead to selecting erroneous cosmological models. Moreover, given the fact that RSD are insensitive to the galaxy bias, one can yield tighter constraints on the measurements of the amplitude of the density perturbations in the power spectrum $\sigma_8$.

\section*{Acknowledgements}
We warmly thank Enzo Branchini, Will Percival, and Fabien Lacasa for their careful read of our manuscript, and Luigi Guzzo for valuable discussions. We acknowledge support from the `Departments of Excellence 2018-2022' Grant awarded by the Italian Ministry of Education, University and Research (\textsc{miur}) L.\ 232/2016. SC is supported by \textsc{miur} through Rita Levi Montalcini project `\textsc{prometheus} -- Probing and Relating Observables with Multi-wavelength Experiments To Help Enlightening the Universe's Structure'.




\bibliographystyle{mnras}
\bibliography{paper1ver2} 

\begin{thebibliography}{}
\makeatletter
\relax
\def\mn@urlcharsother{\let\do\@makeother \do\$\do\&\do\#\do\^\do\_\do\%\do\~}
\def\mn@doi{\begingroup\mn@urlcharsother \@ifnextchar [ {\mn@doi@}
  {\mn@doi@[]}}
\def\mn@doi@[#1]#2{\def\@tempa{#1}\ifx\@tempa\@empty \href
  {http://dx.doi.org/#2} {doi:#2}\else \href {http://dx.doi.org/#2} {#1}\fi
  \endgroup}
\def\mn@eprint#1#2{\mn@eprint@#1:#2::\@nil}
\def\mn@eprint@arXiv#1{\href {http://arxiv.org/abs/#1} {{\tt arXiv:#1}}}
\def\mn@eprint@dblp#1{\href {http://dblp.uni-trier.de/rec/bibtex/#1.xml}
  {dblp:#1}}
\def\mn@eprint@#1:#2:#3:#4\@nil{\def\@tempa {#1}\def\@tempb {#2}\def\@tempc
  {#3}\ifx \@tempc \@empty \let \@tempc \@tempb \let \@tempb \@tempa \fi \ifx
  \@tempb \@empty \def\@tempb {arXiv}\fi \@ifundefined
  {mn@eprint@\@tempb}{\@tempb:\@tempc}{\expandafter \expandafter \csname
  mn@eprint@\@tempb\endcsname \expandafter{\@tempc}}}

\bibitem[\protect\citeauthoryear{{Abbott} et~al.,}{{Abbott}
  et~al.}{2018a}]{Abbott:2018ydy}
{Abbott} T.~M.~C.,  et~al., 2018a, \mn@doi [\apjs] {10.3847/1538-4365/aae9f0},
  \href {https://ui.adsabs.harvard.edu/abs/2018ApJS..239...18A} {239, 18}

\bibitem[\protect\citeauthoryear{Abbott et~al.}{Abbott
  et~al.}{2018b}]{Abbott:2017wau}
Abbott T. M.~C.,  et~al., 2018b, \mn@doi [Phys. Rev.]
  {10.1103/PhysRevD.98.043526}, D98, 043526

\bibitem[\protect\citeauthoryear{Abdalla et~al.}{Abdalla
  et~al.}{2015}]{Abdalla2015}
Abdalla F.~B.,  et~al., 2015, PoS, AASKA14, 17

\bibitem[\protect\citeauthoryear{Addison, Huang, Watts, Bennett, Halpern,
  Hinshaw  \& Weiland}{Addison et~al.}{2016}]{Addison2015}
Addison G.~E.,  Huang Y.,  Watts D.~J.,  Bennett C.~L.,  Halpern M.,  Hinshaw
  G.,   Weiland J.~L.,  2016, \mn@doi [Astrophys. J.]
  {10.3847/0004-637X/818/2/132}, 818, 132

\bibitem[\protect\citeauthoryear{Ade et~al.}{Ade et~al.}{2014}]{Ade2013}
Ade P. A.~R.,  et~al., 2014, \mn@doi [Astron. Astrophys.]
  {10.1051/0004-6361/201321529}, 571, A1

\bibitem[\protect\citeauthoryear{Ade et~al.}{Ade et~al.}{2015}]{Ade2015a}
Ade P. A.~R.,  et~al., 2015, \mn@doi [Phys. Rev. Lett.]
  {10.1103/PhysRevLett.114.101301}, 114, 101301

\bibitem[\protect\citeauthoryear{Ade et~al.}{Ade et~al.}{2016}]{Ade2015}
Ade P. A.~R.,  et~al., 2016, \mn@doi [Astron. Astrophys.]
  {10.1051/0004-6361/201525830}, 594, A13

\bibitem[\protect\citeauthoryear{Amendola et~al.}{Amendola
  et~al.}{2013}]{Amendola2013}
Amendola L.,  et~al., 2013, \mn@doi [Living Rev. Rel.] {10.12942/lrr-2013-6},
  16, 6

\bibitem[\protect\citeauthoryear{Amendola et~al.}{Amendola
  et~al.}{2018}]{Amendola2016}
Amendola L.,  et~al., 2018, \mn@doi [Living Rev. Rel.]
  {10.1007/s41114-017-0010-3}, 21, 2

\bibitem[\protect\citeauthoryear{Bacon et~al.}{Bacon et~al.}{2018}]{SKA1_2018}
Bacon D.~J.,  et~al., 2018, Submitted to: Publ. Astron. Soc. Austral.

\bibitem[\protect\citeauthoryear{Balaguera-Antolínez, Bilicki, Branchini  \&
  Postiglione}{Balaguera-Antolínez et~al.}{2018}]{Balaguera-Antolinez2018}
Balaguera-Antolínez A.,  Bilicki M.,  Branchini E.,   Postiglione A.,  2018,
  \mn@doi [Mon. Not. Roy. Astron. Soc.] {10.1093/mnras/sty262}, 476, 1050

\bibitem[\protect\citeauthoryear{Battye, Charnock  \& Moss}{Battye
  et~al.}{2015}]{Battye2015}
Battye R.~A.,  Charnock T.,   Moss A.,  2015, \mn@doi [Phys. Rev.]
  {10.1103/PhysRevD.91.103508}, D91, 103508

\bibitem[\protect\citeauthoryear{Bernstein}{Bernstein}{2009}]{Bernstein2009}
Bernstein G.~M.,  2009, \mn@doi [Astrophys. J.] {10.1088/0004-637X/695/1/652},
  695, 652

\bibitem[\protect\citeauthoryear{Blake, Collister, Lahav  \& Bridle}{Blake
  et~al.}{2007}]{Blake2006}
Blake C.,  Collister A.,  Lahav O.,   Bridle S.,  2007, \mn@doi [Monthly
  Notices of the Royal Astronomical Society]
  {10.1111/j.1365-2966.2006.11263.x}, 374, 1527

\bibitem[\protect\citeauthoryear{Blas, Lesgourgues  \& Tram}{Blas
  et~al.}{2011}]{Blas2011}
Blas D.,  Lesgourgues J.,   Tram T.,  2011, \mn@doi [JCAP]
  {10.1088/1475-7516/2011/07/034}, 1107, 034

\bibitem[\protect\citeauthoryear{Bonvin \& Durrer}{Bonvin \&
  Durrer}{2011}]{BonvinDurrer2011}
Bonvin C.,  Durrer R.,  2011, \mn@doi [Phys. Rev.]
  {10.1103/PhysRevD.84.063505}, D84, 063505

\bibitem[\protect\citeauthoryear{Cacciato, van~den Bosch, More, Mo  \&
  Yang}{Cacciato et~al.}{2013}]{Cacciato2013}
Cacciato M.,  van~den Bosch F.~C.,  More S.,  Mo H.,   Yang X.,  2013, \mn@doi
  [Mon. Not. Roy. Astron. Soc.] {10.1093/mnras/sts525}, 430, 767

\bibitem[\protect\citeauthoryear{Camera, Santos  \& Maartens}{Camera
  et~al.}{2015a}]{Camera:2014bwa}
Camera S.,  Santos M.~G.,   Maartens R.,  2015a, \mn@doi [Mon. Not. Roy.
  Astron. Soc.] {10.1093/mnras/stv040, 10.1093/mnras/stx159}, 448, 1035

\bibitem[\protect\citeauthoryear{Camera, Maartens  \& Santos}{Camera
  et~al.}{2015b}]{CMS2014}
Camera S.,  Maartens R.,   Santos M.~G.,  2015b, \mn@doi [Mon. Not. Roy.
  Astron. Soc.] {10.1093/mnrasl/slv069}, 451, L80

\bibitem[\protect\citeauthoryear{Camera, Harrison, Bonaldi  \& Brown}{Camera
  et~al.}{2017}]{CameraHarrisonBonaldiBrown}
Camera S.,  Harrison I.,  Bonaldi A.,   Brown M.~L.,  2017, \mn@doi [Mon. Not.
  Roy. Astron. Soc.] {10.1093/mnras/stw2688}, 464, 4747

\bibitem[\protect\citeauthoryear{Camera, Fonseca, Maartens  \& Santos}{Camera
  et~al.}{2018}]{Camera:2018jys}
Camera S.,  Fonseca J.,  Maartens R.,   Santos M.~G.,  2018, \mn@doi [Mon. Not.
  Roy. Astron. Soc.] {10.1093/mnras/sty2284}, 481, 1251

\bibitem[\protect\citeauthoryear{Camera, Martinelli  \& Bertacca}{Camera
  et~al.}{2019}]{Camera2017a}
Camera S.,  Martinelli M.,   Bertacca D.,  2019, \mn@doi [Phys. Dark Univ.]
  {10.1016/j.dark.2018.11.008}, 23, 100247

\bibitem[\protect\citeauthoryear{Challinor \& Lewis}{Challinor \&
  Lewis}{2011}]{ChallinorandLewis2011}
Challinor A.,  Lewis A.,  2011, \mn@doi [Phys. Rev.]
  {10.1103/PhysRevD.84.043516}, D84, 043516

\bibitem[\protect\citeauthoryear{Charnock, Battye  \& Moss}{Charnock
  et~al.}{2017}]{Charnock2017}
Charnock T.,  Battye R.~A.,   Moss A.,  2017, \mn@doi [Phys. Rev.]
  {10.1103/PhysRevD.95.123535}, D95, 123535

\bibitem[\protect\citeauthoryear{Chaves-Montero, Hernández-Monteagudo  \&
  Angulo}{Chaves-Montero et~al.}{2018}]{Jonas2018}
Chaves-Montero J.,  Hernández-Monteagudo C.,   Angulo R.~E.,  2018, \mn@doi
  [Monthly Notices of the Royal Astronomical Society] {10.1093/mnras/sty924},
  477, 3892

\bibitem[\protect\citeauthoryear{Crocce, Cabré  \& Gaztañaga}{Crocce
  et~al.}{2011}]{Crocce2010}
Crocce M.,  Cabré A.,   Gaztañaga E.,  2011, \mn@doi [Monthly Notices of the
  Royal Astronomical Society] {10.1111/j.1365-2966.2011.18393.x}, 414, 329

\bibitem[\protect\citeauthoryear{Di~Dio, Montanari, Lesgourgues  \&
  Durrer}{Di~Dio et~al.}{2013}]{DiDio2013}
Di~Dio E.,  Montanari F.,  Lesgourgues J.,   Durrer R.,  2013, \mn@doi [JCAP]
  {10.1088/1475-7516/2013/11/044}, 1311, 044

\bibitem[\protect\citeauthoryear{Durrer}{Durrer}{2008}]{Durrer2008}
Durrer R.,  2008, The Cosmic Microwave Background.
Cambridge University Press, \mn@doi{10.1017/CBO9780511817205}

\bibitem[\protect\citeauthoryear{Durrer}{Durrer}{2015}]{Durrer2015}
Durrer R.,  2015, \mn@doi [Class. Quant. Grav.]
  {10.1088/0264-9381/32/12/124007}, 32, 124007

\bibitem[\protect\citeauthoryear{Eifler, Krause, Schneider  \&
  Honscheid}{Eifler et~al.}{2014}]{Eifler2014}
Eifler T.,  Krause E.,  Schneider P.,   Honscheid K.,  2014, \mn@doi [Mon. Not.
  Roy. Astron. Soc.] {10.1093/mnras/stu251}, 440, 1379

\bibitem[\protect\citeauthoryear{Elvin-Poole et~al.}{Elvin-Poole
  et~al.}{2018}]{Elvin-Poole:2017xsf}
Elvin-Poole J.,  et~al., 2018, \mn@doi [Phys. Rev.]
  {10.1103/PhysRevD.98.042006}, D98, 042006

\bibitem[\protect\citeauthoryear{Feroz, Hobson  \& Bridges}{Feroz
  et~al.}{2009}]{FHB2009}
Feroz F.,  Hobson M.~P.,   Bridges M.,  2009, \mn@doi [Mon. Not. Roy. Astron.
  Soc.] {10.1111/j.1365-2966.2009.14548.x}, 398, 1601

\bibitem[\protect\citeauthoryear{Fonseca, Camera, Santos  \& Maartens}{Fonseca
  et~al.}{2015}]{FonsecaCameraMaartensSantos}
Fonseca J.,  Camera S.,  Santos M.,   Maartens R.,  2015, \mn@doi [Astrophys.
  J.] {10.1088/2041-8205/812/2/L22}, 812, L22

\bibitem[\protect\citeauthoryear{Foreman-Mackey, Hogg, Lang  \&
  Goodman}{Foreman-Mackey et~al.}{2013}]{ForemanMackey2013}
Foreman-Mackey D.,  Hogg D.~W.,  Lang D.,   Goodman J.,  2013, \mn@doi [Publ.
  Astron. Soc. Pac.] {10.1086/670067}, 125, 306

\bibitem[\protect\citeauthoryear{Grasshorn~Gebhardt \&
  Jeong}{Grasshorn~Gebhardt \& Jeong}{2018}]{HSGGDJ2017}
Grasshorn~Gebhardt H.~S.,  Jeong D.,  2018, \mn@doi [Phys. Rev.]
  {10.1103/PhysRevD.97.023504}, D97, 023504

\bibitem[\protect\citeauthoryear{Heavens \& Taylor}{Heavens \&
  Taylor}{1995}]{Heaven1995}
Heavens A.~F.,  Taylor A.~N.,  1995, \mn@doi [Monthly Notices of the Royal
  Astronomical Society] {10.1093/mnras/275.2.483}, 275, 483

\bibitem[\protect\citeauthoryear{Hoyle et~al.}{Hoyle
  et~al.}{2018}]{Hoyle:2017mee}
Hoyle B.,  et~al., 2018, \mn@doi [Mon. Not. Roy. Astron. Soc.]
  {10.1093/mnras/sty957}, 478, 592

\bibitem[\protect\citeauthoryear{Joachimi \& Bridle}{Joachimi \&
  Bridle}{2010}]{JoachimiBridle2010}
Joachimi B.,  Bridle S.~L.,  2010, \mn@doi [Astron. Astrophys.]
  {10.1051/0004-6361/200913657}, 523, A1

\bibitem[\protect\citeauthoryear{Joudaki et~al.}{Joudaki
  et~al.}{2017a}]{Joudaki2016a}
Joudaki S.,  et~al., 2017a, \mn@doi [Mon. Not. Roy. Astron. Soc.]
  {10.1093/mnras/stw2665}, 465, 2033

\bibitem[\protect\citeauthoryear{Joudaki et~al.}{Joudaki
  et~al.}{2017b}]{Joudaki2016b}
Joudaki S.,  et~al., 2017b, \mn@doi [Mon. Not. Roy. Astron. Soc.]
  {10.1093/mnras/stx998}, 471, 1259

\bibitem[\protect\citeauthoryear{Kaiser}{Kaiser}{1987a}]{Kaiser1987}
Kaiser N.,  1987a, \mn@doi [Monthly Notices of the Royal Astronomical Society]
  {10.1093/mnras/227.1.1}, 227, 1

\bibitem[\protect\citeauthoryear{Kaiser}{Kaiser}{1987b}]{Limber1953}
Kaiser N.,  1987b, \mn@doi [Monthly Notices of the Royal Astronomical Society]
  {10.1093/mnras/227.1.1}, 227, 1

\bibitem[\protect\citeauthoryear{Kaiser}{Kaiser}{1992}]{Kaiser1992}
Kaiser N.,  1992, \mn@doi [The Astrophysical Journal] {10.1086/171151}, 388,
  272

\bibitem[\protect\citeauthoryear{Kitching \& Heavens}{Kitching \&
  Heavens}{2011}]{Kitching:2010wa}
Kitching T.~D.,  Heavens A.~F.,  2011, \mn@doi [Mon. Not. Roy. Astron. Soc.]
  {10.1111/j.1365-2966.2011.18369.x}, 413, 2923

\bibitem[\protect\citeauthoryear{Krause \& Eifler}{Krause \&
  Eifler}{2017}]{KrauseEifler2016}
Krause E.,  Eifler T.,  2017, \mn@doi [Mon. Not. Roy. Astron. Soc.]
  {10.1093/mnras/stx1261}, 470, 2100

\bibitem[\protect\citeauthoryear{Kwan et~al.}{Kwan et~al.}{2017}]{Kwan2016}
Kwan J.,  et~al., 2017, \mn@doi [Mon. Not. Roy. Astron. Soc.]
  {10.1093/mnras/stw2464}, 464, 4045

\bibitem[\protect\citeauthoryear{Laureijs et~al.}{Laureijs
  et~al.}{2011}]{Laureijs2011}
Laureijs R.,  et~al., 2011, ESA/SRE(2011)12

\bibitem[\protect\citeauthoryear{Lesgourgues}{Lesgourgues}{2011}]{Lesgourgues2011}
Lesgourgues J.,  2011, CERN-PH-TH/2011-081, LAPTH-009/11

\bibitem[\protect\citeauthoryear{Lewis, Challinor  \& Lasenby}{Lewis
  et~al.}{2000}]{LCL2000}
Lewis A.,  Challinor A.,   Lasenby A.,  2000, \mn@doi [Astrophys. J.]
  {10.1086/309179}, 538, 473

\bibitem[\protect\citeauthoryear{Liu, Ortiz-Vazquez  \& Hill}{Liu
  et~al.}{2016}]{Liu2016}
Liu J.,  Ortiz-Vazquez A.,   Hill J.~C.,  2016, \mn@doi [Phys. Rev.]
  {10.1103/PhysRevD.93.103508}, D93, 103508

\bibitem[\protect\citeauthoryear{LoVerde \& Afshordi}{LoVerde \&
  Afshordi}{2008}]{LoverdeAfshordi2008}
LoVerde M.,  Afshordi N.,  2008, \mn@doi [Phys. Rev.]
  {10.1103/PhysRevD.78.123506}, D78, 123506

\bibitem[\protect\citeauthoryear{Ma, Hu  \& Huterer}{Ma
  et~al.}{2005}]{MaHuHuterer2006}
Ma Z.-M.,  Hu W.,   Huterer D.,  2005, \mn@doi [Astrophys. J.]
  {10.1086/497068}, 636, 21

\bibitem[\protect\citeauthoryear{Maartens, Abdalla, Jarvis  \& Santos}{Maartens
  et~al.}{2015}]{Maartens2015}
Maartens R.,  Abdalla F.~B.,  Jarvis M.,   Santos M.~G.,  2015, \mn@doi [PoS]
  {10.22323/1.215.0016}, AASKA14, 016

\bibitem[\protect\citeauthoryear{Makarov et~al.,}{Makarov
  et~al.}{2007}]{Padmanabhan2006}
Makarov A.,  et~al., 2007, \mn@doi [Monthly Notices of the Royal Astronomical
  Society] {10.1111/j.1365-2966.2007.11593.x}, 378, 852

\bibitem[\protect\citeauthoryear{Mandelbaum, Slosar, Baldauf, Seljak, Hirata,
  Nakajima, Reyes  \& Smith}{Mandelbaum et~al.}{2013}]{Mandelbaum2013}
Mandelbaum R.,  Slosar A.,  Baldauf T.,  Seljak U.,  Hirata C.~M.,  Nakajima
  R.,  Reyes R.,   Smith R.~E.,  2013, \mn@doi [Mon. Not. Roy. Astron. Soc.]
  {10.1093/mnras/stt572}, 432, 1544

\bibitem[\protect\citeauthoryear{McDonald \& Seljak}{McDonald \&
  Seljak}{2009}]{MacDonaldSelijak2009}
McDonald P.,  Seljak U.,  2009, \mn@doi [JCAP] {10.1088/1475-7516/2009/10/007},
  0910, 007

\bibitem[\protect\citeauthoryear{Nicola, Refregier  \& Amara}{Nicola
  et~al.}{2016}]{NicolaRefregierAmara2016}
Nicola A.,  Refregier A.,   Amara A.,  2016, \mn@doi [Phys. Rev.]
  {10.1103/PhysRevD.94.083517}, D94, 083517

\bibitem[\protect\citeauthoryear{Pourtsidou \& Tram}{Pourtsidou \&
  Tram}{2016}]{PourtsidouTram2016}
Pourtsidou A.,  Tram T.,  2016, \mn@doi [Phys. Rev.]
  {10.1103/PhysRevD.94.043518}, D94, 043518

\bibitem[\protect\citeauthoryear{Raveri}{Raveri}{2016}]{Raveri2016}
Raveri M.,  2016, \mn@doi [Phys. Rev.] {10.1103/PhysRevD.93.043522}, D93,
  043522

\bibitem[\protect\citeauthoryear{Ross, Percival  \& Nock}{Ross
  et~al.}{2010}]{Nock2010}
Ross A.~J.,  Percival W.~J.,   Nock K.,  2010, \mn@doi [Monthly Notices of the
  Royal Astronomical Society] {10.1111/j.1365-2966.2010.16927.x}, 407, 520

\bibitem[\protect\citeauthoryear{Scharf, Fisher  \& Lahav}{Scharf
  et~al.}{1994}]{Fisher1993}
Scharf C.~A.,  Fisher K.~B.,   Lahav O.,  1994, \mn@doi [Monthly Notices of the
  Royal Astronomical Society] {10.1093/mnras/266.1.219}, 266, 219

\bibitem[\protect\citeauthoryear{Seljak}{Seljak}{2009}]{Selijak2009}
Seljak U. c.~v.,  2009, \mn@doi [Phys. Rev. Lett.]
  {10.1103/PhysRevLett.102.021302}, 102, 021302

\bibitem[\protect\citeauthoryear{Singh, Mandelbaum  \& Brownstein}{Singh
  et~al.}{2017}]{Singh2016}
Singh S.,  Mandelbaum R.,   Brownstein J.~R.,  2017, \mn@doi [Mon. Not. Roy.
  Astron. Soc.] {10.1093/mnras/stw2482}, 464, 2120

\bibitem[\protect\citeauthoryear{Spergel, Flauger  \& Hložek}{Spergel
  et~al.}{2015}]{Spergel2015}
Spergel D.~N.,  Flauger R.,   Hložek R.,  2015, \mn@doi [Phys. Rev.]
  {10.1103/PhysRevD.91.023518}, D91, 023518

\bibitem[\protect\citeauthoryear{Szalay, Matsubara  \& Landy}{Szalay
  et~al.}{1998}]{Szalay1998}
Szalay A.~S.,  Matsubara T.,   Landy S.~D.,  1998, \mn@doi [The Astrophysical
  Journal] {10.1086/311293}, 498, L1

\bibitem[\protect\citeauthoryear{Tegmark, Taylor  \& Heavens}{Tegmark
  et~al.}{1997}]{Tegmark:1996bz}
Tegmark M.,  Taylor A.,   Heavens A.,  1997, \mn@doi [Astrophys. J.]
  {10.1086/303939}, 480, 22

\bibitem[\protect\citeauthoryear{Weinberg, Mortonson, Eisenstein, Hirata, Riess
   \& Rozo}{Weinberg et~al.}{2013}]{Weinberg2013}
Weinberg D.~H.,  Mortonson M.~J.,  Eisenstein D.~J.,  Hirata C.,  Riess A.~G.,
   Rozo E.,  2013, \mn@doi [Phys. Rept.] {10.1016/j.physrep.2013.05.001}, 530,
  87

\bibitem[\protect\citeauthoryear{Yahya, Bull, Santos, Silva, Maartens, Okouma
  \& Bassett}{Yahya et~al.}{2015}]{Yahya}
Yahya S.,  Bull P.,  Santos M.~G.,  Silva M.,  Maartens R.,  Okouma P.,
  Bassett B.,  2015, \mn@doi [Mon. Not. Roy. Astron. Soc.]
  {10.1093/mnras/stv695}, 450, 2251

\bibitem[\protect\citeauthoryear{Yoo}{Yoo}{2010}]{Yoo2010}
Yoo J.,  2010, \mn@doi [Phys. Rev.] {10.1103/PhysRevD.82.083508}, D82, 083508

\bibitem[\protect\citeauthoryear{Yoo \& Seljak}{Yoo \&
  Seljak}{2012}]{YooSelijak2012}
Yoo J.,  Seljak U.,  2012, \mn@doi [Phys. Rev.] {10.1103/PhysRevD.86.083504},
  D86, 083504

\bibitem[\protect\citeauthoryear{Zuntz et~al.,}{Zuntz et~al.}{2015}]{Zuntz2015}
Zuntz J.,  et~al., 2015, \mn@doi [Astron. Comput.]
  {10.1016/j.ascom.2015.05.005}, 12, 45

\makeatother
\end{thebibliography}

\appendix

\section{Derivation of Eq.~(10)}
\label{sec:apa1}
We apply the recurrence relations for the spherical Bessel functions to express $j^{\prime\prime}_\ell(k\chi)$ in terms of $j$ functions at different multipoles \citep[see e.g.][]{HSGGDJ2017}. Thence, we obtain \begin{equation}
C^{\rm g,den+RSD}_{\ell\gg1}(z_i,z_j)=\int\frac{\de\chi}{\chi^2}\,\sum_AK^{ij}_A(\chi)P_{\rm lin}\left(k=\frac{\ell+A}{\chi}\right),
\label{eq:CldenRSD_Limber_v1}
\end{equation}
where $A$ is an index that can only take values $1/2$, $-3/2$, or $5/2$, and $K^{ij}_A(\chi)$ is the kernel related to the redshift bin pair $i-j$. We have
\begin{multline}
K^{ij}_{1/2}(\chi)=a_0W_b^i(\chi)W_b^j(\chi)+a_1W_f^i(\chi)W_f^j(\chi)+a_2W_b^i(\chi)W_f^j(\chi)+a_3W_f^i(\chi)W_b^j(\chi)+a_4W_b^i(\chi)W_f^j\left(\frac{\ell-3/2}{\ell+1/2}\chi\right)\\
+a_5W_b^i(\chi)W_f^j\left(\frac{\ell+5/2}{\ell+1/2}\chi\right)+a_6W_f^i(\chi)W_f^j\left(\frac{\ell-3/2}{\ell+1/2}\chi\right)+a_7W_f^i(\chi)W_f^j\left(\frac{\ell+5/2}{\ell+1/2}\chi\right),\label{EQ:KIJ1}
\end{multline}
where we recognise the first term as that in \autoref{eq:Clden_Limber}; this implies $a_0=1$. Then,
\begin{equation}
K^{ij}_{-3/2}(\chi)=a_8W_f^i(\chi)W_b^j\left(\frac{\ell+1/2}{\ell-3/2}\chi\right)+a_9W_f^i(\chi)W_f^j(\chi)+a_{10}W_f^i(\chi)W_f^j\left(\frac{\ell+1/2}{\ell-3/2}\chi\right)+a_{11}W_f^i(\chi)W_f^j\left(\frac{\ell+5/2}{\ell-3/2}\chi\right),\label{EQ:KIJ2}
\end{equation}
and finally,
\begin{equation}
K^{ij}_{5/2}(\chi)=a_{12}W_f^i(\chi)W_b^j\left(\frac{\ell+1/2}{\ell+5/2}\chi\right)+a_{13}W_f^i(\chi)W_f^j(\chi)+a_{14}W_f^i(\chi)W_f^j\left(\frac{\ell+1/2}{\ell+5/2}\chi\right)+a_{15}W_f^i(\chi)W_f^j\left(\frac{\ell-3/2}{\ell+5/2}\chi\right),\label{EQ:KIJ3}
\end{equation}
The coefficients $a_i$ are presented in \autoref{sec:coeffs}. Now, if we perform a change of variable $\tilde\chi=[(\ell+A)/(\ell+1/2)]\chi$, \autoref{eq:CldenRSD_Limber_v1} can be further simplified so that only the usual Limber identity $k=(\ell+1/2)/\chi$ appears. Thus, we have
\begin{multline}
C^{\rm g,den+RSD}_{\ell\gg1}(z_i,z_j)=\int\frac{\de\chi}{\chi^2}\,P_{\rm lin}\left(k=\frac{\ell+1/2}{\chi}\right)\Bigg[\tilde a_0W_b^i(\chi)W_b^j(\chi)+\tilde a_1W_f^i(\chi)W_f^j(\chi)+\tilde a_2W_b^i(\chi)W_f^j(\chi)+\tilde a_3W_f^i(\chi)W_b^j(\chi)\\
+\tilde a_4W_b^i(\chi)W_f^j\left(\frac{2\ell-3}{2\ell+1}\chi\right)+\tilde a_5W_b^i(\chi)W_f^j\left(\frac{2\ell+5}{2\ell+1}\chi\right)+\tilde a_6W_f^i(\chi)W_f^j\left(\frac{2\ell-3}{2\ell+1}\chi\right)+\tilde a_7W_f^i(\chi)W_f^j\left(\frac{2\ell+5}{2\ell+1}\chi\right)\\
+\tilde a_8W_f^i\left(\frac{2\ell-3}{2\ell+1}\chi\right)W_b^j(\chi)+\tilde a_9W_f^i\left(\frac{2\ell-3}{2\ell+1}\chi\right)W_f^j\left(\frac{2\ell-3}{2\ell+1}\chi\right)+\tilde a_{10}W_f^i\left(\frac{2\ell-3}{2\ell+1}\chi\right)W_f^j(\chi)+\tilde a_{11}W_f^i\left(\frac{2\ell-3}{2\ell+1}\chi\right)W_f^j\left(\frac{2\ell+5}{2\ell+1}\chi\right)\\
+\tilde a_{12}W_f^i\left(\frac{2\ell+5}{2\ell+1}\chi\right)W_b^j(\chi)+\tilde a_{13}W_f^i\left(\frac{2\ell+5}{2\ell+1}\chi\right)W_f^j\left(\frac{2\ell+5}{2\ell+1}\chi\right)+\tilde a_{14}W_f^i\left(\frac{2\ell+5}{2\ell+1}\chi\right)W_f^j(\chi)+\tilde a_{15}W_f^i\left(\frac{2\ell+5}{2\ell+1}\chi\right)W_f^j\left(\frac{2\ell-3}{2\ell+1}\chi\right)\Bigg],
\label{eq:CldenRSD_Limber_v2}
\end{multline}
where $\{\tilde a_0,\ldots,\tilde a_7\}=\{a_0,\ldots,a_7\}$, $\{\tilde a_8,\ldots,\tilde a_{11}\}=(2\ell+1)/(2\ell-3)\{a_8,\ldots,a_{11}\}$, and $\{\tilde a_{12},\ldots,\tilde a_{15}\}=(2\ell+1)/(2\ell+5)\{a_{12},\ldots,a_{15}\}$. Eventually, by defining the global den+RSD window function $W^i(\chi)$ of \autoref{eq:W_tot},
we can recast \autoref{eq:CldenRSD_Limber_v2} in the more compact form of \autoref{eq:CldenRSD_Limber}.

\subsection{}
\label{sec:coeffs}
\begin{align}
a_0&=1,\\
a_1&=\left[\frac{2\ell^2+2\ell-1}{(2\ell-1)(2\ell+3)}\right]^2,\\
a_2&=\sqrt{a_1},\\
a_3&=a_2,\\
a_4&=-\frac{\ell(\ell-1)}{(2\ell-1)\sqrt{(2\ell+1)(2\ell-3)}},\\
a_5&=-\frac{(\ell+1)(\ell+2)}{(2\ell+3)\sqrt{(2\ell+1)(2\ell+5)}},\\
a_6&=a_2a_4,\\
a_7&=a_2a_5,\\
a_8&=\frac{2\ell-3}{2\ell+1}a_4,\\
a_9&=\left[\frac{\ell(\ell-1)}{(2\ell-1)(2\ell+1)}\right]^2,\\
a_{10}&=a_2a_8,\\
a_{11}&=\frac{2\ell-3}{\sqrt{(2\ell+5)(2\ell-3)}}\sqrt{a_9a_{13}},\\
a_{12}&=\frac{2\ell+5}{2\ell+1}a_5,\\
a_{13}&=\left[\frac{(\ell+1)(\ell+2)}{(2\ell+1)(2\ell+3)}\right]^2,\\
a_{14}&=a_2a_{12},\\
a_{15}&=\frac{2\ell+5}{2\ell-1}\frac{a_{11}a_4}{a_8}
	\label{tab:table2}
\end{align}




\bsp	
\label{lastpage}
\end{document}